\documentclass[11pt]{article}
\usepackage{graphicx}

\usepackage{geometry}
\usepackage{setspace}
\usepackage{graphicx}
\usepackage{amssymb}
\usepackage{epstopdf}
\usepackage{multirow}
\usepackage{extarrows}
\usepackage{rotating}
\usepackage{dsfont}
\usepackage{bm}
\usepackage{empheq}
\usepackage{endnotes}
\usepackage{tikz}
\newcommand\mathC{\mkern1mu\raise2.2pt\hbox{$\scriptscriptstyle|$}
        {\mkern-7mu\rm C}}              

\title{Qualitative Individuation  in permutation-invariant quantum mechanics} 
           \author{Adam Caulton\footnote{adam.caulton@gmail.com}}
	             
\begin{document}
\maketitle

\begin{abstract}
In this article I expound an understanding of the quantum mechanics of so-called ``indistinguishable'' systems in which permutation invariance is taken as a symmetry of a special kind, namely the result of representational redundancy.  This understanding has heterodox consequences for the understanding of the states of constituent systems in an assembly and for the notion of entanglement.  It corrects widespread misconceptions about the inter-theoretic relations between quantum mechanics and both classical particle mechanics and quantum field theory.

The most striking of the heterodox consequences are: (i) that fermionic states ought not always to be considered entangled; (ii) it is possible for two fermions or two bosons to be discerned using purely monadic quantities; and that (iii)  fermions (but not bosons) may \emph{always} be so discerned.

After an introduction to the key ideas of permutation invariance (Section \ref{Introduction}), I outline and reject in Section \ref{AgainstFac} a widespread interpretation of the quantum formalism associated with ``indistinguishable'' systems.  In Section \ref{QI} I offer my own alternative interpretation, in which quantum systems may be ``qualitatively individuated'', that is, individuated according to their states.  This gives rise to a novel (albeit elementary) technical machinery, which I extend in Section \ref{QIind} to individual systems considered in isolation.  There I give a general prescription for calculating the state of any individual ``indistinguishable'' system.  In Section \ref{Entangle} I explore the implications for the notion of \emph{entanglement} and link up with some recent physics literature on the matter.  In Section \ref{Fermions} I explore the implications for the discernibility of so-called ``indistinguishable'' systems---I will argue that such systems are in fact usually discernible.   I conclude in Section \ref{Conclusion}.
\end{abstract}

\newpage

\tableofcontents

\newpage

\section{Introduction}\label{Introduction}

What would it mean for a quantum mechanical theory to be permutation invariant?  By now the philosophy literature on permutation invariance and related issues in quantum mechanics is formidable, and a variety of construals of permutation invariance have been well articulated.\footnote{A short reading list would have to include: Margenau (1944), French \& Redhead (1988), Butterfield (1993), Huggett (1999), Massimi (2001), Huggett (2003),  Saunders (2003a, 2003b, 2006), French \& Krause (2006, ch.~4), Muller \& Saunders (2008), Muller \& Seevinck (2009), Huggett \& Imbo (2009), Caulton \& Butterfield (2012b), Caulton (2013), Huggett and Norton (2013), Ladyman \emph{et al} (2013). \label{fn:factorists}}  The purpose of this article is to expound and advocate a construal according to which permutation invariance is treated as what I have elsewhere called an \emph{analytic symmetry}.  In analogy with analytic propositions, the truth of analytic symmetries---i.e.~the holding of such symmetries---is a logical consequence of our choice of representational system.  One surely uncontroversial example is the gauge symmetry of electromagnetic theories, which is taken by (almost\footnote{Belot (1998) and Healey (2007) have both entertained denying that electromagnetic $U(1)$ gauge symmetry is ``analytic'', in the face of unpalatable consequences otherwise.}) all to be constituted by a representational redundancy---idle wheels, or ``descriptive fluff'' in the words of Earman (2004)---in the mathematical formalism of the theory.  My motivating idea here is that permutation symmetry is constituted by a similar representational redundancy: quantum mechanics is permutation invariant because what is permuted in the mathematical formalism has no physical reality.

This doctrine, that permutation symmetry is due to representational redundancy, may be familiar, but its unassailable consequences are much less so.  A central negative claim of this paper (in section \ref{AgainstFac}) will be that a great many formal and informal discussions, in both the physics and philosophy of physics literature, tacitly rely on an interpretative doctrine that is mistaken---even explicitly disavowed in those very same discussions.  And no surprise: the doctrine, like many aspects of linguistic practice, is hard to expunge. 

This interpretative doctrine---which I call \emph{factorism}---treats the factor Hilbert spaces of a joint Hilbert space, the latter of which represents the possible states of an assembly of systems, as having separate physical significance, even when permutation invariance has been imposed.  This doctrine justifies, and is required to justify, a number of formal procedures and technical definitions.  There are two particularly important examples: the partial tracing procedure for calculating the states of constituent systems and the definition of entanglement as non-separability.
\begin{enumerate}
\item \emph{Partial tracing.}  Given any state---a density operator---of an assembly of systems, the state of each constituent system is taken to be given  by a reduced density operator obtained by performing a partial trace over all factor Hilbert spaces except the space taken to correspond to the system of interest (see e.g. Nielsen \& Chuang (2010, 105ff.)).

For the purposes of illustration, take for example the joint Hilbert space $\mathfrak{H}:=\mathcal{H}\otimes\mathcal{H}$, and any density operator $\rho \in\mathcal{D}(\mathfrak{H})$ in the class of unit-trace, positive operators on $\mathfrak{H}$.  The reduced state of each of the two constituent systems is then supposed to be given by a partial trace of $\rho$ over the other factor Hilbert space:
\begin{equation}\label{PartialTrace}
\rho_1 = \mbox{Tr}_2(\rho); \qquad \rho_2 = \mbox{Tr}_1(\rho)
\end{equation}

When permutation invariance is imposed, and the relevant joint Hilbert space becomes either the symmetric or anti-symmetric subspace of $\mathfrak{H}$, this procedure  \emph{continues} to be used to extract the states of the constituent systems.

But if permutation invariance is an analytic symmetry, then factor Hilbert space labels represent \emph{nothing} physical; so the prescriptions given in (\ref{PartialTrace}) cannot have anything other than formal significance. (I will later argue that they both yield the ``average state'' of the constituent systems, under an appropriately modified understanding of what the constituent systems are, given in sections \ref{QI} and \ref{QIind}.)

If that is right, then we are in need of an alternative prescription, once permutation invariance has been imposed.  I will suggest an alternative below (section \ref{RDOs}).  An important interpretative upshot is that the claim, commonly found in the philosophy literature, that when permutation invariance is imposed, all constituent systems occupy the same state (usually an improper mixture: always so in the case of fermions), is \emph{false}.  In fact we will see (in section \ref{FermionsAD}) that fermions are \emph{always} in different states, as per the informal and much maligned understanding of Pauli exclusion, and that, in all but a minority of states, so are bosons.

The immediate consequence of this is that fermions obey a certain strong version of Leibniz's Principle of the Identity of Indiscernibles, according to which fermions are always discernible by monadic predicates alone, a.k.a.~being \emph{absolutely discernible}.\footnote{Weyl (1928, 247) seems to have recognised this; he referred to the Pauli exclusion principle as the `Leibniz-Pauli principle'.  Muller and Saunders (2008, 501)  have claimed Weyl (1928) as an early advocate for their own view, based on a principle of charity and the belief that their account is the only one that makes fermions discernible in any non-trivial sense.  But it is doubtful that their subtle notion of weak discernibility is what Weyl had in mind; I contend that it is more plausible that Weyl took fermions to be discernible in the stronger sense defended here.}  As for bosons, we will see (in section \ref{bosons}) that they too are often discernible in the same strong sense, but also that there may be utterly \emph{in}discernible---that is, not even \emph{weakly} discernible---despite recent claims to the contrary (Muller and Seevinck (2009), Caulton (2013), Huggett and Norton (2013)).

\item \emph{Entanglement as non-separability.}
Entanglement for an assembly of systems is almost exclusively defined formally as the non-separability of the assembly's joint state, i.e.~the inability to write it as a product state in any single-system basis (throughout I restrict to pure states of the assembly).  If factorism is correct, then this is indeed justified: for then it would be true to say that a non-separable state represents a joint state of the assembly whose constituent systems do not themselves possess definite (i.e.~pure) states.

But a consequence of the physical emptiness of the factor Hilbert space labels is that we cannot read facts about a physical state so readily off the mathematical form of the state-vector used to represent it.  Instead we need to look at the \emph{algebra} of admissible operators, which is greatly restricted under permutation invariance.\footnote{I am indebted to John Earman for this insight.}  There we find that the sort of entanglement that arises when this restricted algebra operates is much harder to achieve than non-separability: that is, non-separability does not entail entanglement.

I said that entanglement is \emph{almost} exclusively defined as  non-separability.  In fact there are dissenters, chiefly Ghirardi, Marinatto \& Weber (2002) (see also Ghirardi \& Marinatto (2003, 2004, 2005)) (and inspired by them, Ladyman \emph{et al} (2013)), Schliemann \emph{et al} (2001) and Eckert \emph{et al} (2002), all of whom are motivated by considerations that have also inspired this paper. As I will argue below (in section \ref{Entangle}) the notion of entanglement that is most appropriate under permutation invariance agrees with the notions suggested by these authors when they agree with each other.  And I use considerations from the restricted joint algebra to arbitrate between them when they disagree.   A consequence of these considerations will be that fermions are \emph{not} always entangled, inextricably and over cosmic distances, despite commonplace claims to the contrary.

\end{enumerate}

At this point, some readers may be feeling somewhat sceptical about the current project.  Interpretative philosophy of physics may seem like a recherch\'e enterprise at the best of times. But in this case, you may be asking yourself, must more ink be spilt trying to interpret a theoretical framework---elementary many-system quantum mechanics---when we already have a better theory---namely, quantum field theory (henceforth, QFT)---to which to apply our best efforts?  In the state space of quantum field theory, permutation invariance is imposed \emph{at the outset}, so surely there is no danger of being led astray by a redundant formalism.  Why worry about elementary quantum mechanics?

I have two broad answers to this (I concede that they will not satisfy everyone).  The first is that, despite having been superseded, elementary quantum mechanics continues to be invaluable in solving a variety of physical problems---particularly in quantum optics and quantum information theory.  Good conceptual  housekeeping here would mean that we could be confident in physical interpretations without having first to translate everything into the language of Fock space and creation and annihilation operators.  (However, such a translation can be illuminating!  I offer one in section \ref{Fock}.)  And, as I will argue in section \ref{Problems1}, without something like the interpretation I offer here of the quantum formalism, the relation of elementary quantum mechanics to both its successor, QFT, \emph{and} its predecessor, classical particle mechanics, is at best obscure and at worst paradoxical.

Secondly, a better understanding of the framework of many-system elementary quantum mechanics actually \emph{aids} our understanding of the framework of QFT.  Not only will I advocate an interpretation of the quantum formalism that meshes better (than the current orthodoxy) with QFT in the limit of conserved total particle number.  Also, I will argue that the metaphysical conclusions drawn here about elementary quantum mechanics may be exported to QFT. 

I re-iterate that this is predominantly a project of interpretation; my goal is a better understanding of an old theory rather than the generation of new technical results. But interpretation and formal results cannot be sharply separated: interpretation inspires and gives significance to formal results, and formal results offer  precision in the articulation of interpretations.  Thus the novel results in this paper include:
\begin{enumerate}
 \item the articulation of a new method for extracting the states of constituent systems from the state of their assembly (section \ref{QIind});
 \item the definition of an alternative notion of entanglement for bosons and fermions, accompanied by a continuous measure of it (section \ref{Entangle}); and
 \item a proof of the fermionic counterpart to Gisin's Theorem (1991), appropriate for this new notion of entanglement (section \ref{GMW})
\end{enumerate}

I should also mention---although it will already be clear---that my general interpretative approach is a `realist' or `representationalist' one.  That is, I work under the assumption that the quantum formalism represents, or at least \emph{aims} at representing, more or less straightforwardly, underlying physical objects and processes.  This sets me apart from those who advocate a `pragmatic' interpretation of quantum mechanics (such as Healey (2012)), those who advocate a subjective or Bayesian interpretation (e.g. Fuchs (2002)), and those who advocate not attempting an interpretation at all.  

Within the representationalist camp, I also set myself apart from those who deem the quantum formalism to be `incomplete', at least at the microscopic scale; that is, I will not consider any of the various hidden variable solutions, such as the de Broglie-Bohm (``pilot wave'') approach.  That said, my considerations can be taken as limited  to the microscopic realm, so I will not need to address the measurement problem.  The upshot is that the metaphysical claims in this article ought to be acceptable to those who advocate either dynamical collapse, modal, or Everettian (``many world'') approaches.

\subsection{Preliminaries: permutation invariance}\label{prelims}

I will here briefly outline the technicalities that will be in use throughout this paper.  One of the most important mathematical objects will be the single-system Hilbert space, $\mathcal{H}$, and its associated algebra of bounded operators, $\mathcal{B(H)}$.  $\mathcal{H}$ may be any separable Hilbert space. In the taxonomy of Murray and von Neumann (1936), $\mathcal{B(H)}$ and any other algebra encountered in this paper will be of type I. (This will be helpful in section \ref{NatDecomp}, where the commutativity of two single-system algebras will be taken to entail a tensor product structure.)

From the single-system Hilbert space $\mathcal{H}$ we construct the \emph{prima facie} joint Hilbert space for $N$ \emph{equivalent} systems by forming a tensor product:
\begin{equation}\label{HS1}
\mathfrak{H}^{(N)} := \underbrace{\mathcal{H}\otimes\mathcal{H}\otimes\ldots\otimes\mathcal{H}}_{N} \equiv \bigotimes^N\mathcal{H}
\end{equation}
(Fraktur typeface will always be used to denote many-system Hilbert spaces).  The equivalence of our constituent systems is expressed by the fact that each factor Hilbert space in (\ref{HS1}) is a copy of the  same single-system Hilbert space.  Amongst other things, this equivalence means that it makes sense to speak of two constituent systems sharing the same state, so that we may talk e.g.~of multiple occupation of the same single-system state.  

Clearly, the equivalence of two constituent systems in this sense entails that their respective single-system Hilbert spaces, equipped with their operator algebras, are unitarily equivalent.   So if these Hilbert spaces support irreducible representations of some group of spacetime symmetries (e.g.~the Galilei or Poincar\'e group), then the systems must possess the same ``intrinsic'', or state-independent, properties, such as rest mass and intrinsic spin.

The joint Hilbert space $\mathfrak{H}^{(N)}$ supports a unitary representation $P: S_N\to\mathcal{B}(\mathfrak{H}^{(N)})$ of $S_N$, the group of all permutations of $N$ symbols  (for more details, see e.g.~Greiner \& M\"uller (1994, ch.~9) or Tung (1985, ch.~5)).  Let $\{|k\rangle\}$, $k \in \{ 1, 2, \ldots, d:=\mbox{dim}(\mathcal{H})\}$ be an orthonormal basis for the single-system Hilbert space $\mathcal{H}$.  Then $P$  is defined by its action on product states of $\mathfrak{H}^{(N)}$.  Let $k:\{1,2,\ldots N\}\to \{1,2,\ldots d\}$, then for all $\pi \in S_N$,
\begin{equation}
P(\pi)|k(1)\rangle\otimes\ldots\otimes|k(N)\rangle
:=
|(k\circ\pi^{-1})(1)\rangle\otimes\ldots\otimes|(k\circ\pi^{-1})(N)\rangle
\end{equation}
The definition of $P$ is completed by extending by linearity to the whole of $\mathfrak{H}^{(N)}$. 

\emph{Permutation invariance} (henceforth, PI; also known as the \emph{Indistinguishability Postulate}, e.g.~in French \& Krause (2006, 131ff.)) is then an $S_N$-equivariance condition on the joint algebra $\mathcal{B}(\mathfrak{H}^{(N)})$.  That is, for all $\pi\in S_N$ and all $|\psi\rangle\in\mathfrak{H}^{(N)}$,
\begin{equation}
\langle \psi | P^\dag(\pi) QP(\pi)|\psi\rangle = \langle\psi|Q|\psi\rangle
\end{equation} 
In more physical language: PI requires that expectation values for all physical operators (operators that have a physical interpretation) be invariant under permutations of the factor Hilbert spaces. 
The restriction of the joint algebra from $\mathcal{B}(\mathfrak{H}^{(N)})$ to the algebra of operators that satisfy this condition---i.e.,~the $S_N$-equivariant, or \emph{symmetric} operators---constitutes a superselection rule.  The superselection sectors are parameterized by the irreducible representations $\Delta_\mu$  (`irreps') of $S_N$, two of which are one-dimensional: namely, the trivial representation $\Delta_+(\pi)=1$, which corresponds to bosons, and the alternating representation $\Delta_-(\pi) = (-1)^{\#(\pi)}$ (where $\#(\pi)$ is the number of inversions in the permutation $\pi$), which corresponds to fermions.  Let us label the boson sector of $\mathfrak{H}^{(N)}$ $\mathfrak{H}_+^{(N)}$ and the fermion sector $\mathfrak{H}_-^{(N)}$.  (If $N\geqslant 3$, then there are also multi-dimensional irreps, corresponding to various types of paraparticle; so $\mathfrak{H}_+^{(N)}\oplus \mathfrak{H}_-^{(N)} \sqsubset \mathfrak{H}^{(N)}$, where `$\sqsubset$' denotes proper subspacehood. Paraparticles will not be considered for the rest of this paper.)  

A note about terminology: I will talk about assemblies of ``indistinguishable'' systems whenever PI is imposed (I use scare quotes, because my main contention is of course  that the systems \emph{are} distinguishable!) and talk about ``distinguishable'' systems whenever PI is imposed---even if the systems are equivalent in the sense above, namely that they are individually represented by copies of the same Hilbert space.

Let me say a few words on the justification of permutation invariance.  The only real justification can be that it leads to empirical adequacy, and that it does is not in question.\footnote{Cf.~e.g.~Prugove\v{c}ki (1981, 307): `It has been observed that the assumption that the state $\Psi$ of the system (e.g.,~gas) $\mathfrak{S}$ in which particle $\mathfrak{S}_k$ is in the state $\Psi_k$ and particle $\mathfrak{S}_l$ in the state $\Psi_l$ is identical to the state of $\mathfrak{S}$ in which $\mathfrak{S}_k$ is in the state $\Psi_l$
and $\mathfrak{S}_l$ in the state $\Psi_k$  leads to a correct energy distribution in the case of particles of integer spin.'}  Alternative justifications of an \emph{a priori} character are sometimes seen in the literature.  One runs along the lines that, since the constituent systems of the assembly possess all the same ``intrinsic'' properties, one would not expect the expectation value of any beable to change upon a permutation of those systems.\footnote{Messiah \& Greenberg (1964, B250), the \emph{locus classicus} for permutation invariance in quantum mechanics, write, `Any one of these permutations [$\pi\in S_N$] is a mere reshuffling of the labels attached to the particles belonging to the same species.  Since these particles are identical, it must not lead to any observable effects.'  Similar claims are made in the textbooks, e.g.~Rae (1992, 205): `Identical particles are often referred to as \emph{indistinguishable} in order to emphasize the fact that they cannot be distinguished by any physical measurement.  This implies that an operator representing any physical measurement on the system must remain unchanged if the labels assigned to the individual particles are interchanged.'}  This justification is suspect for at least two reasons.  First, it presumes a representational connection between the mathematical formalism and the physics which it is my main interest here to deny; namely that factor Hilbert space labels represent or denote particles (for it is the factor Hilbert space labels that are  being permuted in PI).

Second, even granted this representational connection, the inference from identical ``intrinsic'' properties to permutation-invariant expectation values for all beables is invalid.  At first blush, nothing seems particularly suspect about the beable, `location of system 17' (represented by the operator $\bigotimes^{16}\mathds{1}\otimes \mathbf{Q}\otimes\bigotimes^{N-17}\mathds{1}$, where $\mathbf{Q}$ is the single-system position operator), and one would certainly \emph{not} expect this beable to preserve its expectation value upon a change of the label `17' to, say, `5', since then one would be talking about the location of system 5, and there is no reason to think that, just because systems 5 and 17 share the same ``intrinsic'' properties, it follows that they must share the same location.\footnote{Huggett and Imbo (2009) give an excellent critique of this fallacious reasoning.}

It would be a mistake to reply to this that PI does nothing but  express the truism that we could have labelled system 17 e.g. with the numeral `5' instead.\footnote{E.g. Merzbacher (1997, 535): `Since the order in which the particles are labeled has, by the definition of particle identity, no physical significance, state vectors (or wavefunctions) that differ only in the permutation of the labels must define the same state.'} PI cannot express that truism, first of all because it entails non-trivial empirical predictions, such as the divergence of assemblies in thermal states from classical Maxwell-Boltzmann statistics---how could such a non-trivial, contingent fact be explained by the truism that we could have chosen a different notational convention from the one that we in fact use?  And secondly, PI does not express that truism because it is reflected instead in the possibility that, for any state-vector $|\psi\rangle$ in $\mathfrak{H}^{(N)}$, we could instead have used $P(5, 17)|\psi\rangle$ to represent the same physical state that we currently use $|\psi\rangle$ to represent.  And all that this requires is that the two algebras associated with each labelling convention be unitarily equivalent. But that is trivial, since $P(5,17)\mathcal{B}(\mathfrak{H}^{(N)})P^\dag(5,17)$ (where $(5,17)$ is the permutation that swaps symbols 5 and 17) and $\mathcal{B}(\mathfrak{H}^{(N)})$ are not only unitarily equivalent, they are identical.

There is another justification for PI that is  more credible.  This justification is that factor Hilbert space labels (i.e.~the order in which the factor Hilbert spaces lie in the tensor product) represent \emph{nothing at all}.  \emph{If} they represented nothing at all, then we certainly \emph{would} expect---indeed, we would have to \emph{ensure}---that beables' expectation values were invariant under their permutation.  Otherwise a permutation would represent a physical difference, and that would contradict the original claim that the things being permuted represented nothing.

In fact the only problem with this justification is that, for it to be useful \emph{as a justification}, we need already to have come to the conclusion that factor Hilbert space labels represent nothing.  How do we come to that conclusion?  Well, the representational connections between our mathematical formalism and the physical world are a matter of our choosing, but on that choice depends the experimental predictions one draws from the theory, i.e.~what we take our theory to be saying.  (It might even be better to say that theories are distinguished not only by their formalisms but also by the representational connections that we determine for them.)  And PI, or at least its consequences for the collective behaviour of particles, is an example of those predictions.  So to justify PI in this way, we already need to know that it is obeyed---but that is just our original justification: empirical adequacy.

While this means that we cannot take our doctrine about Hilbert space labels to justify PI, perhaps we can reverse the situation by instead taking the empirical success of PI as abductive support for that doctrine.  This would require the absence of anything physical, to which  the factor Hilbert space labels would otherwise correspond, to be the best explanation for PI.  Is that the best explanation of PI?  I see no reason to think that the case  here is any worse than the analogous case in electromagnetism, in which gauge invariance is best explained by the physical unreality of the four-vector potential field.

Other reasons, if not to support the doctrine that factor Hilbert space labels represent nothing, then at least to \emph{renounce} the doctrine that factor Hilbert space labels represent or denote the constituent systems of the assembly, are given in sections \ref{Problems1} and \ref{Problems2}.  But first I will say a little more about that doctrine.

\section{Against factorism}\label{AgainstFac}

This section is dedicated to dispensing with the doctrine I call factorism.  In section \ref{facmdefined}, I give a precise definition of this doctrine. In section \ref{facmisnothaecm}, I contrast factorism with haecceitism, a more familiar interpretative doctrine in the quantum philosophy literature.  My criticisms of factorism comprise two sections: section \ref{Problems1} discusses the trouble factorism causes for the inter-theoretic relation between quantum mechanics and both classical particle mechanics and QFT; and section \ref{Problems2} argues that factorism contravenes the principle that unitary equivalent representation represent the same physical possibilities.

\subsection{Factorism defined}\label{facmdefined}

Factorism says: particles are the physical correlates of the labels of factor Hilbert spaces.  This view is orthodox: indeed, not just orthodox, but  well-nigh universal.\footnote{It is a key interpretative assumption in {all} of the articles mentioned in footnote \ref{fn:factorists},  except for Ladyman \emph{et al} (2013).} It is deeply entrenched in the way we all speak and think, and learn, about quantum mechanics for more than one system. To explain this, and how the view is nonetheless deniable, it will be clearest to begin by considering first, an assembly of  equivalent but ``distinguishable''  systems (in the sense of section \ref{prelims}).  Here the relevant joint Hilbert space is $\mathfrak{H}^{(N)}$.

Each factor Hilbert space $\mathcal{H}$ in the expression (\ref{HS1}) for $\mathfrak{H}^{(N)}$ represents the space of pure states for each system.  The full space of states---including the mixed states---for each particle is then represented by $\mathcal{D(H})$, the space of density operators defined on $\mathcal{H}$.  Factorism now takes the position that the $i$th copy of $\mathcal{H}$ ($\mathcal{D(H})$) represents  the possible pure (mixed) states for the $i$th system, so that each Hilbert space label---i.e.~its position in the tensor product in (\ref{HS1})---may be taken to represent or denote its corresponding system.

I concur. I am happy to take this step: I agree that ``distinguishable'' systems \emph{are} the physical correlates of the labels of factor Hilbert spaces, in the usual tensor-product formalism.

But factorism goes beyond this agreement. It says that the same goes  for ``indistinguishable'' systems: i.e.~that the labels of the factor Hilbert spaces represent or denote their corresponding systems. Or in other words: \emph{although} under PI such an assembly is described by the symmetric ($\mathfrak{H}^{(N)}_+$) or antisymmetric  ($\mathfrak{H}^{(N)}_-$) subspace of the tensor product  space $\mathfrak{H}^{(N)}$, this does \emph{not} disrupt the factor spaces' labels referring to the constituent systems. Thus when one treats an assembly using the symmetric or antisymmetric subspace of  ($\mathfrak{H}^{(N)}$), factorism says that there are $N$ constituent systems, one for each factor space, and the $i$th system's pure (mixed) states are represented by the state-vectors in the $i$th copy of $\mathcal{H}$ (density operators in the $i$th copy of  $\mathcal{D(H})$).

\subsection{Factorism and haecceitism}\label{facmisnothaecm}

Some readers may be wondering how the doctrine I have called factorism differs from {\em haecceitism}, an interpretative doctrine which is more familiar in the philosophy of quantum theory (and logic and metaphysics quite generally).  In fact they are quite different---indeed they are logically independent, as I shall argue below.

Haecceitism and its denial, anti-haecceitism, are instances of a general issue to do with the  representation of possibilities. In philosophers' jargon, the issue is whether a distinction is real, as against `merely verbal', `spurious' or a `distinction without a difference'.  In the context of mathematical physics and its interpretation, the issue comes down to whether there are \emph{redundancies} in the representation of  physical possibilities by the mathematical objects in our theory's formalism.  That is, whether the representation relation between mathematical states (vectors or rays of the Hilbert space) and physical states is one-to-one or many-to-one.

The specific distinction with which haecceitism is associated concerns, like factorism, the action of permutations on states. There may be some mathematical states $\rho$ (which in our case are the rays or minimal projectors of the joint Hilbert space) that are wholly symmetric in the sense that their orbit under this action is a singleton set, i.e. contains only the state in question: $S_N(\rho) = \{\rho\}$. But typically a generic state $\rho$ will have a non-singleton orbit.  So the question arises whether all the elements of the orbit $S_N(\rho)$ represent the same \emph{physical} state of affairs. 

{\em Anti-haecceitism} may be defined as always answering `Yes' to this question (cf.~Lewis (1986, 221)). This answer implements the intuitive idea of treating the underlying individuality of each system, the `which-is-which-ness' of the systems, as physically empty or unreal.  {\em Haecceitism} therefore says, on the contrary, that distinct mathematical states in an orbit represent distinct physical states.  Intuitively, this implements the idea that the underlying individuality of the systems is real. However, strictly speaking if haecceitism is just the denial of anti-haecceitism, then it need not be committed to such specific claims; it need only \emph{deny} the rather global claim that permuted mathematical states always correspond to the same physical state.

Haecceitism, in the sense just defined, is general enough to be assessed in classical or in quantum mechanics.  Finite-dimensional classical mechanics (e.g.~of $N$ point particles) is, so far as I know, almost always formulated haecceitistically, i.e.~so as to distinguish states differing by a permutation of indistinguishable particles (although Belot (2001, pp.~56-61) considers the anti-haecceitistic alternative). And in infinite-dimensional classical mechanics (i.e.,~the mechanics of continuous  media---fluids or solids), one \emph{must} be a haecceitist (Butterfield (2011, pp.~358-61)).

In quantum mechanics under PI, the assessment of haecceitism is trivialized by the fact that every state of $\mathfrak{H}^{(N)}_+$ and $\mathfrak{H}^{(N)}_-$ is fixed by all permutations, i.e.~$P(\pi)|\psi\rangle\langle\psi|P^\dag(\pi) = |\psi\rangle\langle\psi|$, so we never get an orbit of permuted states larger than a singleton set.  (See Pooley (2006).  The situation changes when we consider paraparticles; see Caulton and Butterfield (2012b)). This already shows that factorism and haecceitism are not the same doctrine; for while the issue of haecceitism vs.~its denial cannot even be articulated for ``indistinguishable'' systems, the issue of factorism vs.~\emph{its} denial can be articulated---and answered.  

At this point I must raise a problem with the definition of `haecceitism' that I just gave.  Consider: if the precise definition of haecceitism in terms of system permutations is to be true to the metaphysical spirit of haecceitism, we must assume that the system labels must represent the objects whose which-is-which-ness the haecceitist wants to defend as real.  But that assumption is just factorism.  Even the standard understanding of haecceitism in quantum mechanics presumes factorism!

But a more noncommittal definition of `haecceitism' is easy to formulate.  We retain the emphasis on system permutations, but refrain from assuming that system permutations are represented by permutations of the factor Hilbert space.  That is, we refrain from interpreting the unitary representation $P$ of $S_N$ as permuting constituent systems.  (We do not give an \emph{alternative} interpretation; we simply refrain from giving any interpretation.)

We therefore have \emph{two} formulations of haecceitism: a general formulation, which talks of generating physical differences by a permutation of \emph{whatever represents or denotes the constituent systems}, and a specific formulation, which talks of generating physical differences by a permutation of factor Hilbert spaces.  Only if factorism is true are these two formulations equivalent.

We just saw that if factorism {is} right (and we ignore paraparticles), then the question of haecceitism cannot be settled.  If, as I believe, factorism is wrong, then the specific formulation (now stripped of its standard metaphysical interpretation) remains irresolvable.  There are then two remaining possibilities for the truth-value of the general formulation of haecceitism.  Both are consistent, but anti-haecceitism seems to be favoured.  This is because, in each of the joint Hilbert spaces $\mathfrak{H}^{(N)}_+$ and $\mathfrak{H}^{(N)}_-$, states identified by specifying occupation numbers for single-system states appear \emph{at most once}.  It follows that the mathematical structure required to even \emph{define} a permutation of systems does not exist.  Thus haecceitism (in the broad sense) could be maintained by an anti-factorist only if she is prepared to declare the standard mathematical formalism incomplete: a coherent move, but a dubious one.

In summary, factorism is a proposal about how the constituent systems of an assembly are represented in the formalism; while haecceitism, in its broad formulation, is a doctrine about how the individuality or `which-is-which-ness' of those systems contributes to the individuation of joint states.  Neither entails the other, but if factorism is wrong, then it is best also to consider haecceitism false.

\subsection{Problems with inter-theoretic relations}\label{Problems1}

Now I will turn to the first of my two criticisms of factorism, which is that it causes trouble both for obtaining a classical limit and for viewing elementary quantum mechanics as a limit of quantum field theory, when total particle number is conserved.  This criticism relies on an inescapable consequence of permutation invariance and factorism: namely, that all constituent systems in an assembly possess the \emph{same} state.

This can also be expressed in terms of the reduced density operators of the constituent systems.  According to the usual procedure of yielding the reduced density operator of a particle by tracing out the states for all the other particles in the assembly (see e.g.~Nielsen \& Chuang (2010, 105)), we obtain the result that for all (anti-) symmetrized joint states, one obtains \emph{equal} reduced density operators for every system.\footnote{A selected bibliography for this result runs as follows: Margenau (1944), French \& Redhead (1988), Butterfield (1993), Huggett (1999, 2003), Massimi (2001), French \& Krause (2006, pp.~150-73).  All these authors interpret this result as showing that Leibniz's Principle of the Identity of Indiscernibles is pandemically {\em false} in quantum theory.}  Moreover, unless the joint state is bosonic, and a product of identical factors (e.g.~$|\phi\rangle\otimes|\phi\rangle\otimes\ldots\otimes|\phi\rangle$), then the reduced state of every system will be statistically mixed.

One consequence of the ensuing ``non-individuality'' of factorist systems is that they cannot  become classical particles in an appropriate limit.  This phenomenon is well discussed by Dieks and Lubberdink (2011), but to summarise: In the classical limit, factorist systems do not even approximately acquire the trajectories we associate with classical particles, since the former must remain in statistically mixed states all the way to the classical limit, or else possibly (if they are bosons) \emph{all} remain in the same pure state.  Factorist systems cannot tend, in any limit, to become distinguished one from another in space---like classical particles---by zero or at least negligible overlap, since each factorist system always possesses the ``entire spatial profile'' of the assembly.  

A similar point can be made about quantum field theory: factorist particles do not tend  to the behaviour of QFT-quanta if we consider the limit in which the total particle number in conserved.  This is because QFT-quanta---which are associated with creation and annihiliation operators $a^\dag(\phi), a(\phi)$, for some state $\phi$ in the single-particle Hilbert space $\mathcal{H}$---always occupy pure states.  Indeed: it may be shown (though I will not here) that QFT-quanta behave just like classical particles in an appropriate classical limit for QFT.  Thus the factorist systems emerge as the embarrassing odd ones out.  
Something has gone wrong.

\subsection{Problems with unitary equivalence}\label{Problems2}

My second criticism of factorism is that it defies an interpretative principle that ought to be compulsory; namely that the unitary equivalence of two Hilbert spaces and accompanying algebras is a sufficient condition for considering those Hilbert spaces to be equally good mathematical representations of the same space of physical possibilities.

Here we run into a potentially confusing ambiguity, which is an occupational hazard of doing interpretative philosophy of physics: the philosophical term `representation', in the sense of a mathematical formalism representing physical facts, must be distinguished from the technical term `representation', in the sense of a map from an abstract algebra to a concrete algebra of operators defined on some Hilbert space.  To resolve this ambiguity, I will use the term `rep' for the technical concept, and continue to use `representation' for the philosophical one.

More specifically, let us define a \emph{rep} of a $*$-algebra $\mathfrak{A}$ as an ordered pair $\langle \mathfrak{H}, \pi\rangle$ of a Hilbert space $\mathfrak{H}$ and a $*$-homomorphism $\pi: \mathfrak{A} \to\mathcal{B}(\mathfrak{H})$.  Then two {reps} $\langle \mathfrak{H}, \pi\rangle$ and $\langle \mathfrak{K}, \phi\rangle$ are  \emph{unitarily equivalent} iff there is a unitary operator $U: \mathfrak{H} \to \mathfrak{K}$ such that $U\pi[\mathfrak{A}] = \phi[\mathfrak{A}]U$ (i.e.~there is an \emph{intertwiner} $U$ between $\pi[\mathfrak{A}]$ and $\phi[\mathfrak{A}]$).

I emphasise that the unitary equivalence of two Hilbert spaces is a much stronger condition than their being isomorphic (which requires only that the have the same dimension). Unitary equivalence means that the \emph{same} abstract algebra of operators is being realised by the concrete algebras of two Hilbert spaces in such a way that preserves \emph{all} expectation values.  It seems uncontroversial, then, that we should consider any two unitarily equivalent reps to represent the same space of physical states, and furthermore we should consider the intertwiner $U$ to preserve physical interpretations.

However, there is a wrinkle here: not every physically relevant quantity may be represented by the abstract algebra.  If mathematical artefacts of one rep but not another are also doing representational work, then we have a reason to block the inference from unitary equivalence by the intertwiner $U$ to preserved physical interpretation under action by $U$.

For example, consider the joint Hilbert space $\mathfrak{H}^{(2)} := \mathcal{H}\otimes\mathcal{H}$, and suppose that the two systems they describe are distinguished one from the other by at least one ``intrinsic'' or state-independent property $F$ (perhaps a haecceity) that one of the systems---but not the other---possesses.  Suppose that it is the \emph{first} copy of $\mathcal{H}$ in the tensor product that represents the system with $F$.
How could $F$ be represented in the mathematical formalism?  Certainly not by any non-trivial operator on $\mathcal{H}$, since the property is state-independent.  But not by a trivial operator---some multiple of the identity, $\lambda\mathds{1}$, where $\lambda\in\mathbb{C}$---on $\mathcal{H}$ either, since \emph{both} systems are represented by that Hilbert space, and it would yield the same value both times.   Similar considerations lead to the conclusion that there is no way to represent $F$ on $\mathfrak{H}^{(2)}$ either.

The only solution is to treat one of the factor Hilbert spaces as distinguished, as it were ``outside'' the formalism.  But the honour of being distinguished in this way will not be preserved under unitary equivalence.  For example, select some subspace $\mathcal{S}_1$ of the first copy of $\mathcal{H}$ (the copy that represents the system with $F$) and some subspace $\mathcal{S}_2$ of the second copy.  Then the joint Hilbert space $\mathfrak{H}_s := \mathcal{S}_1\otimes\mathcal{S}_2$ represents a subspace of the joint Hilbert space $\mathfrak{H}^{(2)}$.  But $\mathfrak{H}_s$, with its associated algebra, is unitarily equivalent to $\mathfrak{H}_p := \mathcal{S}_2\otimes\mathcal{S}_1$ (`$p$' for `permute') and its associated algebra: the intertwiner is the permutation operator $P(12)$ restricted to $\mathfrak{H}_s$.  And while it is true that $\mathfrak{H}_p$ would represent equally well the physical states currently represented by $\mathfrak{H}_s$, it is \emph{not} true that $P(12)$ preserves the  physical interpretation of the states between $\mathfrak{H}_s$ and $\mathfrak{H}_p$.  For example, the state $|\Psi\rangle:=|\phi\rangle\otimes|\chi\rangle$ represents the physical state in which the system with $F$ is in the state (represented by) $|\phi\rangle$ and the system without $F$ is in the state (represented by) $|\chi\rangle$.  But the permuted state $P(12)|\Psi\rangle$ instead represents the physical state in which the system with $F$ is in the state (represented by) $|\chi\rangle$ and the system without $F$ is in the state (represented by) $|\phi\rangle$.
In summary, so long as there are deemed to be \emph{any} physical quantities not represented in the relevant algebras, we cannot allow that the intertwiner between two unitary equivalent reps preserves the physical interpretation of the components of those two reps.

The remedy for this wrinkle is simply to add a invariance condition to our  interpretative principle:
\begin{quote}
If two {reps} are unitarily equivalent, \emph{and any physical quantity not represented in either algebra is invariant between them}, then the intertwiner between the two reps preserves the physical interpretation of any component of those two reps.
\end{quote}
This deals with the problem above, since the identity of the system that possesses $F$ is not invariant between $\mathfrak{H}_s$ and $\mathfrak{H}_p$.

Factorism runs afoul of this interpretative principle.  For, as the results of the next section (\ref{NatDecomp}) show, there are subspaces of the symmetric and anti-symmetric spaces $\mathfrak{H}^{(N)}_+$ and $\mathfrak{H}^{(N)}_-$ that are unitarily equivalent to subspaces of $\mathfrak{H}^{(N)}$.  And so long as we take the order of the factor Hilbert spaces to have no physical significance (for this is not invariant between the subspaces), we ought, by the principle above, to give the same physical interpretation to these subspaces.  But that requires denying the result mentioned above in section \ref{Problems1} that, under  permutation invariance, all constituent systems are in the same physical state.  And the only way to resist that result is to relinquish the interpretative doctrine that justifies taking the partial trace of a joint state to obtain the reduced state of its constituents, which is factorism.

\section{Qualitative individuation}\label{QI}

This section inaugurates the positive part of this paper.  The foregoing criticisms of factorism have left us in need of an alternative procedure for extracting the states of constituent systems from the joint state of an assembly.  What we need is some way to decompose the joint Hilbert space ($\mathfrak{H}^{(N)}_+$ or $\mathfrak{H}^{(N)}_-$) in a way that obeys the strictest interpretation of permutation invariance. 

\subsection{Decomposing the right joint Hilbert space}\label{DecompRight}

One might say that factorism fails by trying to decompose the wrong joint Hilbert space.  Rather than decomposing $\mathfrak{H}^{(N)}_+$ or $\mathfrak{H}^{(N)}_-$ into single-system Hilbert spaces, it instead forgets the superselection rule induced by permutation invariance, embeds $\mathcal{H}^{(N)}_+$ and $\mathfrak{H}^{(N)}_-$ back into $\mathfrak{H}^{(N)}$, and proceeds to decompose that instead.  And the decomposition of this space is easy, since by construction it has a tensor product structure.  However, what we want to do is take the superselection rule seriously, and try to decompose $\mathfrak{H}^{(N)}_+$ or $\mathfrak{H}^{(N)}_-$.

A natural idea is to try to find a similar $N$-fold tensor product structure in $\mathfrak{H}^{(N)}_+$ or $\mathfrak{H}^{(N)}_-$.  In more detail, this would entail  finding $N$ putatively single-system Hilbert spaces $\mathfrak{h}_1, \mathfrak{h}_2, \ldots \mathfrak{h}_N$, with associated single-system algebras, such that the tensor product Hilbert space formed from these, $\mathfrak{h}_1\otimes \mathfrak{h}_2\otimes\ldots\otimes \mathfrak{h}_N$, is unitarily equivalent to one of $\mathfrak{H}^{(N)}_+$ or $\mathfrak{H}^{(N)}_-$.

 But we face an immediate problem: these joint spaces may well have a \emph{prime} number of dimensions,\footnote{For example, if $\mathcal{H} \cong \mathbb{C}^2$, then $\mathfrak{H}^{(2)}_+ \cong \mathbb{C}^3$.}  which entails that only one of the would-be factor Hilbert spaces $\mathfrak{h}_k$ could have more than 1 dimension!  Worse: it is hard to see what physical interpretation, in terms of single-system states, one could give to the multi-dimensional factor space $\mathfrak{h}_k$.

However, this doomed idea can be rehabilitated: for we need not be so ambitious so as to decompose the \emph{whole} of the joint Hilbert space in one go.  Instead, we can try decomposing \emph{subspaces} of $\mathfrak{H}^{(N)}_+$ or  $\mathfrak{H}^{(N)}_-$.  
 
Suppose such a decomposition successful in princple for some subspace $\mathfrak{S}$.  Then the collection of constituents corresponding to the decomposition must be interpreted as co-existing \emph{only} in those states belonging to $\mathfrak{S}$.  This is not objectionable \emph{per se}.  Agreed: in the case of ``distinguishable systems'' there are means of individuating systems which will suffice for all states.  But if one is not a haecceitist, why should one demand or expect this all the time?

If there are two subspaces, $\mathfrak{S}_1$ and $\mathfrak{S}_2$ say, each of which may be decomposed, then the question arises whether any constituent system represented in $\mathfrak{S}_1$ is the same as any constituent system represented in $\mathfrak{S}_2$.  It will turn out that this question does not have an unequivocal answer.  The associated metaphysical picture is one in which relations of ``trans-state identity'' (as we might call it) have no objective significance, a quantum analogue of Lewis's (1968) celebrated Counterpart Theory.  We should not be surprised: as we saw in section \ref{facmisnothaecm}, the anti-factorist picture, like Lewis's, eschews haecceitism.

Now that we have limited our search for decompositions to subspaces of the joint Hilbert space, it remains to be shown that such subspaces exist.  It is the purpose of the next section to prove that they do.

\subsection{Natural decompositions}\label{NatDecomp}

I will use the term \emph{individuation} for the act of picking out a object, or collection of objects, according to some property that it may have; and I will call the property in question the \emph{individuation criterion}.  Since individuation in this sense need not entail uniqueness, an individuation criterion is not quite a Russellian definite description. 

I will also  say that an object, or class of objects, is \emph{qualitatively} individuated iff its individuation criterion is a qualitative property.  I will assume that qualitative properties are represented by projectors in the single-system Hilbert space $\mathcal{H}$. (Factorist individuation---i.e.~individuation according to factor Hilbert space labels---may be considered non-qualitative individuation.) These projectors need not be minimal, i.e.~1-dimensional.  But a minimal projector corresponds to a maximally logically strong qualitative property (something like a Leibnizian `individual concept').

Since I deny factorism,  in the context of PI qualitative individuation is our only means of picking out constituent systems.  I will argue here that the decomposable {subspaces} of the joint Hilbert space may correspond to physical states in which the constituent systems have been qualitatively individuated.

What counts as a successful decomposition of a joint Hilbert space?  What are our criteria for success?  Here I draw upon the work of Zanardi (2001) and Zanardi \emph{et al} (2004), which emphasises working in terms of algebras of beables.  Zanardi (2001, p.~3) writes:
\begin{quote}
When is it legitimate to consider a pair of observable algebras as describing a bipartite quantum system? Suppose that $\mathcal{A}_1$ and $\mathcal{A}_2$ are two \emph{commuting} $*$-subalgebras of $\mathcal{A} := \mbox{End}(\mathfrak{H})$ such that the subalgebra $\mathcal{A}_1\vee\mathcal{A}_2$ they generate, i.e.,~the minimal $*$-subalgebra containing both $\mathcal{A}_1$ and $\mathcal{A}_2$, amounts to the whole $\mathcal{A}$, and moreover one has the (noncanonical) algebra isomorphism, 
\begin{equation}\label{Zanardi}
\mathcal{A}_1\vee\mathcal{A}_2 \cong \mathcal{A}_1\otimes\mathcal{A}_2
\end{equation}
The standard, \emph{genuinely} bipartite, situation is of course $\mathfrak{H} = \mathcal{H}_1\otimes\mathcal{H}_2, \mathcal{A}_1 = \mbox{End}(\mathcal{H}_1)\otimes\mathds{1}$, and $\mathcal{A}_2 = \mathds{1}\otimes\mbox{End}(\mathcal{H}_2)$.  If $\mathcal{A}'_i  := \{X\ |\ [X,\mathcal{A}_1]=0 \}$ denotes the \emph{commutant} of $\mathcal{A}_1$, in this case one has $\mathcal{A}'_i = \mathcal{A}_2$.
\end{quote}
Thus Zanardi's proposal is to work by analogy with the case of distinguishable systems: we look for commuting subalgebras whose tensor product is isomorphic to the joint (symmetric) algebra for the assembly's Hilbert space.  The `(noncanonical) algebra isomorphism' can for us be unitary equivalence.

Zanardi \emph{et al} (2004, p.~1) offer three necessary and jointly sufficient conditions for what they coin a \emph{natural decomposition}.  I will express these by setting $N=2$, for definiteness (so that we are seeking a decomposition into two constituent systems).
\begin{itemize}
\item \emph{Local accessibility.}  This condition states that the subalgebras be `controllable', i.e.~experimentally implementable.  We have no need of this condition here, since our interest is not experimental but metaphysical.  But we may replace it with the condition that the subalgebras have a natural physical interpretation as \emph{single}-system algebras.  This means that the beables in these algebras ought to have a recognizable form as monadic quantities, such as position, momentum and spin.
\item \emph{Subsystem independence.}  This condition requires the subalgebras to {commute}: 
\begin{equation}
\forall A\in\mathcal{A}_1, \forall B\in\mathcal{A}_2: [A, B]= 0.
\end{equation}
I.e., each system ought to possesses its properties \emph{independently} of the other. This condition will be familiar from algebraic quantum field theory, where it is imposed on observable algebras associated with space-like separated, compact spacetime regions under the name \emph{microcausality}, and interpreted as vetoing the possibility of act-outcome correlations between space-like separated events. (For more details, see Halvorson (2007, sections 2.1, 7.2).)  
\item \emph{Completeness.}  This condition requires the minimum algebra containing both subalgebras to amount to the original joint algebra $\mathcal{A}$:
\begin{equation}
\mathcal{A} = \mathcal{A}_1\vee\mathcal{A}_2
\end{equation}
This expresses the fact that the assembly in question has been decomposed \emph{without residue}.
\end{itemize} 

We are here dealing only with separable Hilbert spaces, and therefore, in the taxonomy of Murray and von Neumann (1936), with type I algebras; it follows that the second two conditions entail the tensor product structure expressed above in (\ref{Zanardi}).  And we read the isomorphism `$\cong$' as a claim of unitary equivalence.  Thus we can say our joint Hilbert space and its associated algebra has been naturally decomposed iff it is unitarily equivalent to the tensor product of identifiably single-system Hilbert spaces and their associated algebras.

\subsection{Individuation blocks}\label{IndBlocks}

I will now show that qualitatively individuated systems provide the natural decompositions being sought.

Let us consider what algebra of operators we may associate with a single, qualitatively individuated system (for simplicity I will concentrate on the two-system case).  Recall that qualitative individuation is individuation by projectors.  So suppose that our two individuation criteria (one for each of the two systems) are given by the projectors $E_\alpha, E_\beta$, each of which acts on the single-system Hilbert space $\mathcal{H}$.

It is important that $E_\alpha \perp E_\beta$, i.e.~$E_\alpha E_\beta = E_\beta E_\alpha = 0$, so that none of the two systems is individuated by the other's criterion.  (The importance of this condition will soon become clear.)  Call the system individuated by $E_\alpha$ the $\alpha$-system, and the system individuated by $E_\beta$, the $\beta$-system.

Now define the following projector on $\mathfrak{H}^{(2)}$:
\begin{equation}\label{QIsubspace}
\mathcal{E} := E_\alpha\otimes E_\beta + E_\beta\otimes E_\alpha
\end{equation}
$\mathcal{E}$ is symmetric, i.e.~it satisfies PI, and its range has non-zero components in both the symmetric ($\mathfrak{H}_+^{(2)}$) and anti-symmetric ($\mathfrak{H}^{(2)}_-$) joint Hilbert spaces.  

I now claim that both $\mathcal{E}\left[\mathfrak{H}^{(2)}_+\right]$ (bosonic) and $\mathcal{E}\left[\mathfrak{H}^{(2)}_-\right]$ (fermionic) have natural decompositions into the single-system spaces 
 $E_\alpha[\mathcal{H}]$ and  $E_\beta[\mathcal{H}]$.  I call these subspaces \emph{individuation blocks}.

To prove this, we need to establish both that: (i) the single-system spaces have an identifiable single-system interpretation; and (ii) the joint Hilbert spaces $\mathcal{E}\left[\mathfrak{H}^{(2)}_+\right]$  and $\mathcal{E}\left[\mathfrak{H}^{(2)}_-\right]$ are unitary equivalent to  $E_\alpha[\mathcal{H}]\otimes E_\beta[\mathcal{H}]$.

The first condition is satisfied, since $E_\alpha[\mathcal{H}]$ and $E_\beta[\mathcal{H}]$ are just subspaces of the familiar single-system Hilbert space $\mathcal{H}$, and we may preserve all physical interpretations under the projections by the individuation criteria $E_\alpha$ and $E_\beta$. 

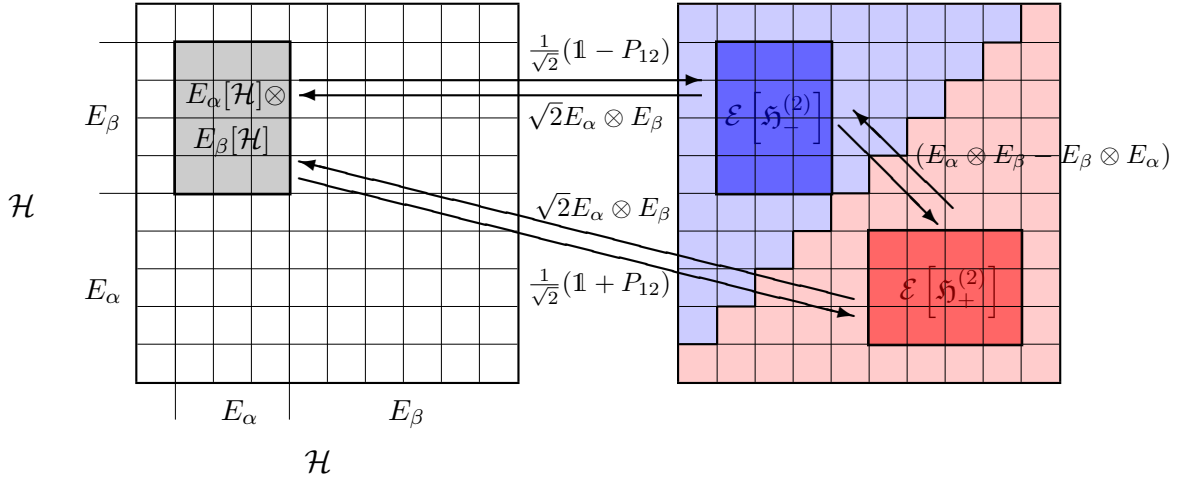
\begin{figure}
\setlength{\unitlength}{1mm}  
\begin{picture}(150,60)   
 %
        \definecolor{grey}{rgb}{.8, .8, .8}
          \definecolor{fermions}{rgb}{.8, .8, 1}
        \definecolor{bosons}{rgb}{1, .8, .8}
          \definecolor{fermionsgrey}{rgb}{.4, .4, 1}
        \definecolor{bosonsgrey}{rgb}{1, .4, .4}
        \definecolor{fermiontext}{rgb}{0,0,.5}
        \definecolor{bosontext}{rgb}{.5,0,0}
\linethickness{15mm}
   \color{grey}
 \put(17.5,35){\line(0,1){20}}   
  \color{black}    
 \thicklines
\put(5,10){\framebox(50,50)}
\linethickness{0.05mm}
 \put(10,10){\line(0,1){50}}
 \put(15,10){\line(0,1){50}}
 \put(20,10){\line(0,1){50}}
 \put(25,10){\line(0,1){50}}
 \put(30,10){\line(0,1){50}}
 \put(35,10){\line(0,1){50}}
  \put(40,10){\line(0,1){50}}
 \put(45,10){\line(0,1){50}}
  \put(50,10){\line(0,1){50}}
  \put(5,15){\line(1,0){50}}
  \put(5,20){\line(1,0){50}}
  \put(5,25){\line(1,0){50}}
  \put(5,30){\line(1,0){50}}
  \put(5,35){\line(1,0){50}}
  \put(5,40){\line(1,0){50}}
  \put(5,45){\line(1,0){50}}
  \put(5,50){\line(1,0){50}}
   \put(5,55){\line(1,0){50}}
  \linethickness{0.1mm}%
 \put(10,5){\line(0,1){50}}
 \put(25,5){\line(0,1){50}}
   \put(0,55){\line(1,0){25}}
  \put(0,35){\line(1,0){25}}
\thicklines%
\put(10,35){\framebox(15,20)}
\put(-2,21){$E_\alpha$}\put(16,5){$E_\alpha$}
\put(38,5){$E_\beta$} \put(-2,44){$E_\beta$} 
\put(27,-2){\large $\mathcal{H}$}
\put(-12,32){\large $\mathcal{H}$}
 \linethickness{50mm}
 \color{bosons}
\put(75,35){\line(1,0){50}}  
\linethickness{5mm}%
\color{fermions}
\put(75,17.5){\line(1,0){5}}  
\put(75,22.5){\line(1,0){10}}  
\put(75,27.5){\line(1,0){15}}  
\put(75,32.5){\line(1,0){20}}  
\put(75,37.5){\line(1,0){25}}  
\put(75,42.5){\line(1,0){30}}  
\put(75,47.5){\line(1,0){35}} 
\put(75,52.5){\line(1,0){40}}   
\put(75,57.5){\line(1,0){45}}  
\linethickness{15mm}%
\color{fermionsgrey}%
 \put(87.5,35){\line(0,1){20}} 
   \color{bosonsgrey}
\put(100,22.5){\line(1,0){20}}   
  \color{black}    %
 \thicklines%
\put(75,10){\framebox(50,50)}
\linethickness{0.05mm}
 \put(80,10){\line(0,1){50}}
 \put(85,10){\line(0,1){50}}
 \put(90,10){\line(0,1){50}}
 \put(95,10){\line(0,1){50}}
 \put(100,10){\line(0,1){50}}
 \put(105,10){\line(0,1){50}}
  \put(110,10){\line(0,1){50}}
 \put(115,10){\line(0,1){50}}
  \put(120,10){\line(0,1){50}}
  \put(75,15){\line(1,0){50}}
  \put(75,20){\line(1,0){50}}
  \put(75,25){\line(1,0){50}}
  \put(75,30){\line(1,0){50}}
  \put(75,35){\line(1,0){50}}
  \put(75,40){\line(1,0){50}}
  \put(75,45){\line(1,0){50}}
  \put(75,50){\line(1,0){50}}
   \put(75,55){\line(1,0){50}}
 \thicklines%
\put(100,15){\framebox(20,15)}
\put(80,35){\framebox(15,20)}
\put(75,15){\line(1,0){5}}
\put(80,20){\line(1,0){5}}
\put(85,25){\line(1,0){5}}
\put(90,30){\line(1,0){5}}
\put(95,35){\line(1,0){5}}
\put(100,40){\line(1,0){5}}
\put(105,45){\line(1,0){5}}
\put(110,50){\line(1,0){5}}
\put(115,55){\line(1,0){5}}
\put(80,15){\line(0,1){5}}
\put(85,20){\line(0,1){5}}
\put(90,25){\line(0,1){5}}
\put(95,30){\line(0,1){5}}
\put(100,35){\line(0,1){5}}
\put(105,40){\line(0,1){5}}
\put(110,45){\line(0,1){5}}
\put(115,50){\line(0,1){5}}
\put(120,55){\line(0,1){5}}
\put(10,47){$E_\alpha[\mathcal{H}] \otimes$}
 \put(11,41){$E_\beta[\mathcal{H}]$}
\color{fermiontext}
\put(81,44){$\mathcal{E}\left[\mathfrak{H}^{(2)}_-\right]$}
\color{bosontext}
\put(104,21){$\mathcal{E}\left[\mathfrak{H}^{(2)}_+\right]$}
\color{black}
\thicklines%
\put(25,50){\vector(1,0){53}}
\put(25,37){\vector(4,-1){73}}
\put(96,44){\vector(1,-1){13}}
\put(78,48){\vector(-1,0){53}}
\put(98,21){\vector(-4,1){73}}
\put(111,33){\vector(-1,1){13}}
\put(55,53){\small $\frac{1}{\sqrt{2}}(\mathds{1} - P_{12})$}
\put(55,22){\small $\frac{1}{\sqrt{2}}(\mathds{1} + P_{12})$}
\put(106,39){\small $(E_\alpha\otimes E_\beta - E_\beta\otimes E_\alpha)$}
\put(55,44){\small $\sqrt{2}E_\alpha\otimes E_\beta$}
\put(56,32){\small $\sqrt{2}E_\alpha\otimes E_\beta$}
\end{picture} 
\caption{The intertwiners between the joint Hilbert spaces $\mathcal{E}\left[\mathfrak{H}^{(2)}_+\right]$, $\mathcal{E}\left[\mathfrak{H}^{(2)}_-\right]$ and $E_\alpha[\mathcal{H}]\otimes E_\beta[\mathcal{H}]$.  The square on the left represents $\mathfrak{H}^{(2)}$ in some product basis and the square on the right represents $\mathfrak{H}^{(2)}$ in the corresponding ``symmetry basis'', in which all fermion states (blue) and all boson states (red) are grouped together. \label{intertwiners}}
\end{figure}

The second condition is satisfied, since we can provide the explicit forms of the intertwiners between the three joint Hilbert spaces; they are given in Figure \ref{intertwiners}. (It may be helpful to point out that, while these operators are not unitaries on the whole of $\mathfrak{H}^{(2)}$, they are when restricted to their relevant domains \emph{so long as} $E_\alpha$ and $E_\beta$ are orthogonal.)

In a little more detail, select for example any two single-system operators $A$ and $B$ on $\mathcal{H}$.  Then $E_\alpha AE_\alpha\otimes E_\beta BE_\beta$ belongs to the tensor product algebra $\mathcal{B}(E_\alpha[\mathcal{H}]\otimes E_\beta[\mathcal{H}]) \cong \mathcal{B}(E_\alpha[\mathcal{H}])\otimes\mathcal{B}(E_\beta[\mathcal{H}])$, and is sent to
\begin{equation}\label{Op1}
 E_\alpha AE_\alpha\otimes E_\beta BE_\beta + E_\alpha AE_\alpha\otimes E_\beta BE_\beta
\end{equation}
under the intertwiner $U_\pm : E_\alpha[\mathcal{H}]\otimes E_\beta[\mathcal{H}] \to \mathcal{E}\left[\mathfrak{H}^{(2)}_\pm\right]$, where $U_\pm := \frac{1}{\sqrt{2}}(\mathds{1}\pm P(12))$.  $U_\pm$ also clearly sends product states $|\phi\rangle\otimes|\chi\rangle$, where $|\phi\rangle\in E_\alpha[\mathcal{H}], |\chi\rangle\in E_\beta[\mathcal{H}]$ to states of the form $\frac{1}{\sqrt{2}}(|\phi\rangle\otimes|\chi\rangle \pm |\chi\rangle\otimes|\phi\rangle)$.

The foregoing results apply to \emph{any} subspace $\mathcal{E}\left[\mathfrak{H}^{(2)}\right]$, as defined in (\ref{QIsubspace}), so long as $E_\alpha \perp E_\beta$, and we can straightforwardly generalise to assemblies of more than two systems.

A class of instances of qualitative individuation that is of particular interest arises when the single-system Hilbert space $\mathcal{H}$ represents more than one degree of freedom.  In this case, if the individuation criteria $E_\alpha, E_\beta$ apply to less than the full degrees of freedom, then the full algebra of linear bounded operators on the remaining degrees of freedom is available to the qualitatively individuated systems.

\begin{figure}[t]
\centering
\includegraphics[width=\textwidth]{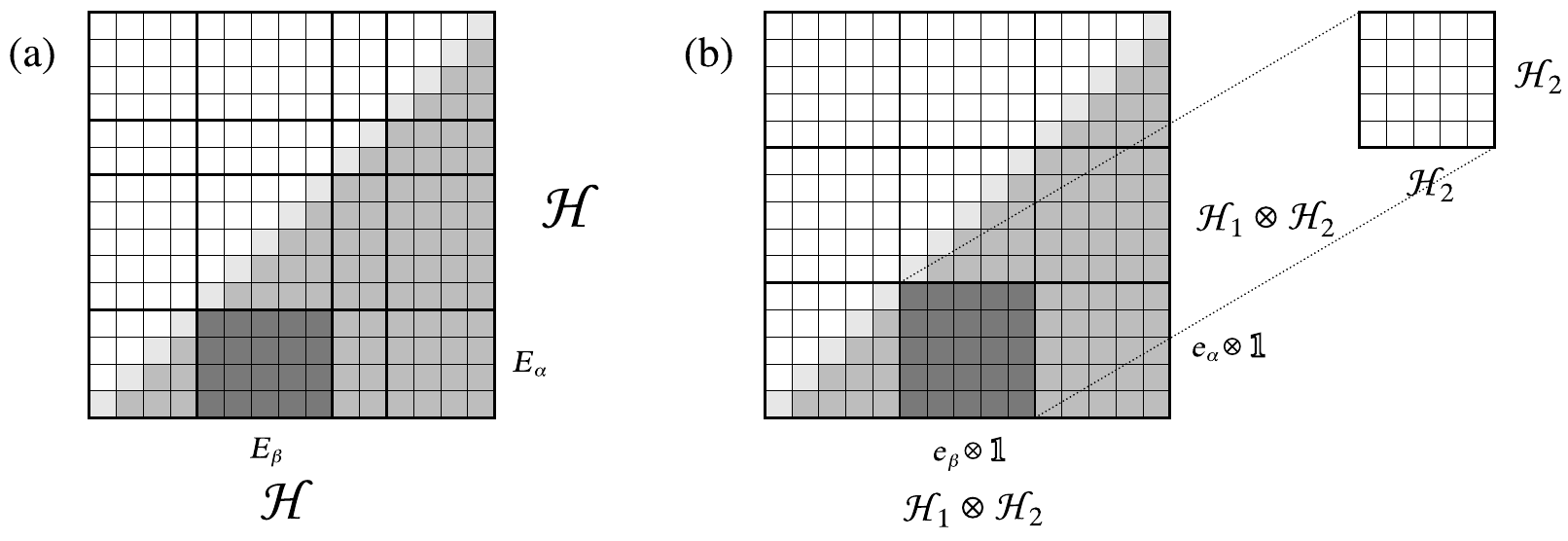}
\caption{(a) The (anti-) symmetric projection of the tensor product of two Hilbert spaces may be decomposed into spaces which exhibit a tensor product structure.   (Light grey squares indicate condensed states, which remain under symmetrization but not anti-symmetrization.) (b) If the two Hilbert spaces are decomposed into eigensubspaces of only one degree of freedom, then the ``off-diagonal" elements of the decomposition serve as irreps for the full algebra of operators for  the other degrees of freedom. \label{QIfig1}}
\end{figure}

For simplicity, suppose that $\mathcal{H}$ represents two degrees of freedom; i.e.,~$\mathcal{H} 
= \mathcal{H}_1 \otimes \mathcal{H}_2$.
Now let us choose the individuation criteria $E_\alpha  = e_\alpha\otimes \mathds{1}, E_\beta = e_\beta\otimes\mathds{1}$, where $e_\alpha$ and $e_\beta$ act on $\mathcal{H}_1$ and $\mathds{1}$ is the identity on $\mathcal{H}_2$.  From the results above, it follows that the algebra of the $\alpha$-system $\mathcal{A}_\alpha = \mathcal{B}(\mbox{ran}(E_\alpha)) = \mathcal{B}(\mbox{ran}(e_\alpha))\otimes\mathcal{B}(\mathcal{H}_2)$ and the algebra of the $\beta$-system is $\mathcal{A}_\beta = \mathcal{B}(\mbox{ran}(E_\beta)) = \mathcal{B}(\mbox{ran}(e_\beta))\otimes\mathcal{B}(\mathcal{H}_2)$.  Thus the full algebra $\mathcal{B}(\mathcal{H}_2)$ is available to both qualitatively individuated systems (cf.~Figure \ref{QIfig1}(b)).\footnote{This case corresponds to Huggett and Imbo's (2009, pp.~315-6) `approximately distinguishable' systems.}

 Finally, we may wish to  consider individuating by multiple individuation blocks.  We may break down the \emph{entire} joint Hilbert space $\mathfrak{H}^{(2)}_\pm$ into those subspaces that do count as individuation blocks and those that do not, and then provide natural decompositions for each of the off-diagonal subspaces.  Each individuation block is associated with its own pair of qualitatively individuated systems, and thus behaves in its own right like a joint Hilbert space representing an assembly of two \emph{distinguishable} systems; cf.~Figure \ref{QIfig1}(a).

In more detail: We may decompose the single-system Hilbert space $\mathcal{H}$ using a complete family of projectors $\{E_i\}, \sum_i E_i = \mathds{1}$: 
\begin{equation}
\mathcal{H} = \left(\sum_i E_i\right)\mathcal{H} = \bigoplus_i E_i[\mathcal{H}] =: \bigoplus_i \mathcal{H}_i
\end{equation}

Then, with $\mathcal{S}_\pm$ the appropriate symmetry projector, the joint Hilbert space is
\begin{eqnarray}
\mathfrak{H}^{(2)}_\pm
&=&
\mathcal{S}_\pm\left[\left(\bigoplus_i \mathcal{H}_i\right)\otimes\left(\bigoplus_i \mathcal{H}_i\right)\right] \\
&=&
\mathcal{S}_\pm\left[\bigoplus_i \left(\mathcal{H}_i\otimes \mathcal{H}_i\right) \oplus \bigoplus_{i<j}\left[(\mathcal{H}_i\otimes \mathcal{H}_j) \oplus (\mathcal{H}_j\otimes \mathcal{H}_i)\right]\right] \\
&=&
\bigoplus_i \mathcal{S}_\pm\left(\mathcal{H}_i\otimes \mathcal{H}_i\right) \oplus \bigoplus_{i<j}\mathcal{S}_\pm\left[(\mathcal{H}_i\otimes \mathcal{H}_j)  \oplus (\mathcal{H}_j\otimes \mathcal{H}_i)\right] \\
&=&
\bigoplus_i \mathcal{S}_\pm\left(\mathcal{H}_i\otimes \mathcal{H}_i\right) \oplus \bigoplus_{i<j} U_\pm\left(\mathcal{H}_i \otimes\mathcal{H}_j\right) U^\dag_\pm  \label{indblocks}
\end{eqnarray}
where $U_\pm$ is the intertwiner defined above.  Each $(i,j)$ term on the RHS of (\ref{indblocks}) corresponds to an individuation block.

One might hope to cover, as far as possible, the entire joint Hilbert space $\mathfrak{H}^{(2)}_\pm$ with individuation blocks, so that one achieves a decomposition of the full joint space after all.  However, these hopes are forlorn, for two reasons.  First, in the case of bosons, multiply occupied states (of the form $|\phi\rangle\otimes|\phi\rangle$) will never lie in any individuation block, since it would require individuation criteria to become non-orthogonal.  These states are always out of reach.  Second, in all cases, the imposition of multiple sets of individuation criteria induces a superselection rule between individuation blocks, since the unitary equivalence results above rely only to each individuation block in isolation.  Therefore one cannot impose multiple sets of individuation criteria without losing information about the joint state.

\subsection{Is Permutation Invariance compulsory?}

Here is an appropriate place to make some  brief comments about Huggett and Imbo's (2009, pp.~313) recent claim that it is not necessary to impose permutation invariance on systems with identical ``intrinsic'' (state-independent) properties.  This is because, they claim, systems may be distinguished according to their `trajectories' (i.e.~single-system states).  If they are correct, this would entail that factorism is, after all, a viable interpretative position for such systems---so long as we understand factor Hilbert space labels as representing these trajectories (just as, in the case of distinguishable systems, we use factor Hilbert space labels to represent distinct state-\emph{independent} properties of the systems).  

The results of this section show that Huggett and Imbo are partly correct.  In my jargon: they are right that an un-symmetrised Hilbert space is an equally adequate (since unitarily equivalent) means to represent an assembly of qualitatively individuated systems---\emph{for those states in which the individuation criteria succeed}; and that therefore there is no \emph{practical} need to impose IP, or, therefore, to repudiate factorism \emph{when represented those states and those states alone}.  But they are wrong to claim that it is not necessary to impose PI to represent \emph{all of the available states} for systems with identical intrinsic properties.  For  the unitary equivalence result above, on which Huggett and Imbo's claim depends, holds \emph{only for the appropriate individuation block}.  As soon as the assembly's state has components that lie outside of this subspace, the equivalence breaks down.

I must emphasise too that the breakdown of unitary equivalence outside of the relevant individuation block does not just mean that, for states outside this subspace, the two formalisms yield conflicting empirical claims (which come down in favour of PI).  Rather, the quasi-factorist formalism ceases to make physical \emph{sense} for states outside of the relevant individuation block.  For, outside of this subspace, the systems no longer occupy the states upon which their individuation---and therefore the entire quasi-factorist formalism---was based.

Huggett and Imbo mistakenly suppose that imposing PI prevents one from qualitatively individuating systems. (As Huggett and Imbo (2009, p.~315) put it: `[PI] $\Rightarrow$ trajectory indistinguishability'.)  But that inference assumes what I deny: namely, factorism.  Without factorism, we can agree with Huggett and Imbo that systems may be qualitatively individuated, \emph{without} contravening PI.  Moreover: without factorism but with PI, we may represent \emph{all} of the states available to the assembly, without fear that our representational apparatus will break down.

\subsection{Russellian vs.~Strawsonian approaches to individuation} \label{QIRvS}

All of the results of the previous sections apply only to joint states that have non-zero support within some individuation block.  This is the subspace for which the  individuation criteria for the systems  fully succeeds; i.e.~for which the projector 
\begin{equation}
\mathcal{E}(\alpha, \beta) := E_\alpha \otimes E_\beta +  E_\beta \otimes E_\alpha
\end{equation}  
has eigenvalue 1.  What about states for which individuation does not succeed?  The question is important, since we want a procedure for calculating expectation values of quantities which belong to the joint algebra of the qualitatively individuated quantities; and we want that procedure to be as general as possible.

The way to proceed depends on one's stance toward reference failure for individuation criteria.  I see two equally acceptable routes, which may be associated (a little tendentiously) with the classic debate over reference failure for definite descriptions.  With a little poetic licence, I call the two routes \emph{Russellian} and \emph{Strawsonian}.

The Russellian route (cf.~Russell 1905) takes the claim of success of the individuation criteria $E_\alpha$ and $E_\beta$ to be an implicit additional claim to any explicit claim which implements those criteria. Thus the expectation value for any  quantity $Q \in \mathcal{B}\left(\mathcal{E}(\alpha, \beta)\left[\mathfrak{H}^{(2)}_\pm\right]\right)$ is given by 
\begin{equation}\label{Russell}
 \langle Q \rangle^{(R)}_{(\alpha, \beta)} := \mbox{Tr}(\mathcal{E} (\alpha, \beta) \rho \mathcal{E}(\alpha, \beta) Q),
  \end{equation}
  which uses the usual quantum  mechanical specifications for expressing conjunction.  But $Q$ commutes with $\mathcal{E}(\alpha, \beta)$, since it is a sum of products of single-system quantities, each of which commutes with $E_\alpha$ and $E_\beta$. So (\ref{Russell}) may be simplified to $\mbox{Tr}(\rho \mathcal{E}(\alpha, \beta)Q)$.

The Strawsonian route (cf.~Strawson 1950) instead takes the joint success of the individuation criteria $E_\alpha$ and $E_\beta$ as a \emph{presupposition}  of any claim which uses that strategy.  Therefore  any expectation values calculated under the presupposition of the success of $\mathcal{E}(\alpha, \beta)$ must be renormalized by conditionalizing on that success.  This is done using the usual L\"uder rule
\begin{equation}\label{Strawson}
\rho \quad\mapsto\quad \rho_{\alpha\beta}:=\frac{\mathcal{E}(\alpha, \beta)\rho\mathcal{E}(\alpha, \beta)}{\mbox{Tr}(\rho\mathcal{E}(\alpha, \beta))}\ .
\end{equation}
The expectation value of any quantity  $Q \in \mathcal{B}\left(\mathcal{E}(\alpha, \beta)\left[\mathfrak{H}^{(2)}_\pm\right]\right)$ is then given simply by $\langle Q \rangle^{(S)}_{(\alpha, \beta)} := \mbox{Tr}(\rho_{\alpha\beta}Q)$. 

Note that conditionalization requires that $\mbox{Tr}(\rho\mathcal{E}(\alpha, \beta)) >0$, which means that the state must have \emph{some} support in the individuation block.  The fact that  $\mbox{Tr}(\rho_{\alpha\beta}Q)$ is undefined when $\mbox{Tr}(\rho \mathcal{E}(\alpha, \beta)) =0$  meshes rather nicely with Strawson's well-known claim that statements containing failed definite descriptions do not possess a truth value.

It will have been noted that the difference between the Russellian and Strawsonian routes for expectation values lies only in the multiplicative factor $\frac{1}{\mbox{\footnotesize Tr}(\rho\mathcal{E}(\alpha, \beta))}$.  A point in favour of the Strawsonian approach is that the identity  $\mathds{1}_{\mathfrak{H}^{(2)}_\pm}$ has expectation value $\langle\mathds{1}_{\mathfrak{H}^{(2)}_\pm}\rangle^{(S)}_{(\alpha, \beta)}=1$ for all normalizable states, while under the Russellian route  the identity's expectation value is equal to $\mathcal{E}(\alpha, \beta)$'s expectation value, which may take any value from 0 to 1.  A point in favour of the Russellian approach is that expectation values may be defined for \emph{all} states.  On this approach, if the assembly's state has \emph{no} support in the individuation block, then expectation value for every quantity  is zero.

\section{Individuating single systems}\label{QIind}

In this section, I turn away from the problem of \emph{completely} decomposing an assembly into constituent systems, and turn instead to the problem of picking out a \emph{single} constituent system from the assembly.  I seek a means to calculate expectation values for quantities associated with a single qualitatively individuated system, whose individuation criterion we may choose.

\subsection{Expectation values for constituent systems}\label{QIexp}

The way I will proceed is inspired in part by the Strawsonian approach to individuation in Section \ref{QIRvS}.  The main idea, there and here, is to \emph{conditionalize} upon the success of the individuation.  As usual, I work, for the sake of simplicity, in the $N=2$ case (unless otherwise stated); the generalization to $N>2$ will be obvious.

\begin{figure}
\setlength{\unitlength}{1mm}  
\begin{center}
\begin{picture}(50,50)   
  \definecolor{fermions}{rgb}{.8, .8, 1}%
  \definecolor{bosons}{rgb}{1, .8, .8}%
\definecolor{fermionsgrey1}{rgb}{.4, .4, 1}%
\definecolor{bosonsgrey1}{rgb}{1, .4, .4}%
\definecolor{fermionsgrey2}{rgb}{0, 0, 1}%
\definecolor{bosonsgrey2}{rgb}{1, 0, 0}%
  \linethickness{50mm}%
  \color{bosons} %
  \put(-4,30){\line(1,0){50}}   %
\linethickness{5mm}%
   \color{fermions} %
\put(-4,12.5){\line(1,0){5}}   
\put(-4,22.5){\line(1,0){15}}   
\put(-4,32.5){\line(1,0){25}}   
\put(-4,42.5){\line(1,0){35}}   
\put(-4,52.5){\line(1,0){45}}   
\put(-4,17.5){\line(1,0){10}}     
\put(-4,27.5){\line(1,0){20}}     
\put(-4,37.5){\line(1,0){30}}     
\put(-4,47.5){\line(1,0){40}}     
\color{fermionsgrey2}%
\put(11,27.5){\line(1,0){5}}
\put(11,32.5){\line(1,0){10}}
\color{bosonsgrey2}%
\put(11,22.5){\line(1,0){15}}
\put(16,27.5){\line(1,0){10}}
\put(21,32.5){\line(1,0){5}}
\linethickness{15mm} %
\color{fermionsgrey1}  %
\put(-5,27.5){\line(1,0){15}}   
\put(17.5,35){\line(0,1){20}}   
\color{bosonsgrey1}  %
\put(25,27.5){\line(1,0){20}}   
\put(17.5,5){\line(0,1){15}}   
  \color{black}   %
 \thicklines  %
\put(-5,5){\framebox(50,50)}
\linethickness{0.05mm}%
 \put(0,5){\line(0,1){50}}
 \put(5,5){\line(0,1){50}}
 \put(10,5){\line(0,1){50}}
 \put(15,5){\line(0,1){50}}
 \put(20,5){\line(0,1){50}}
 \put(25,5){\line(0,1){50}}
  \put(30,5){\line(0,1){50}}
 \put(35,5){\line(0,1){50}}
  \put(40,5){\line(0,1){50}}
  \put(-5,10){\line(1,0){50}}
  \put(-5,15){\line(1,0){50}}
  \put(-5,20){\line(1,0){50}}
  \put(-5,25){\line(1,0){50}}
  \put(-5,30){\line(1,0){50}}
  \put(-5,35){\line(1,0){50}}
  \put(-5,40){\line(1,0){50}}
  \put(-5,45){\line(1,0){50}}
   \put(-5,50){\line(1,0){50}}
 \thicklines%
\put(-5,10){\line(1,0){5}}
\put(0,15){\line(1,0){5}}
\put(5,20){\line(1,0){5}}
\put(10,25){\line(1,0){5}}
\put(15,30){\line(1,0){5}}
\put(20,35){\line(1,0){5}}
\put(25,40){\line(1,0){5}}
\put(30,45){\line(1,0){5}}
\put(35,50){\line(1,0){5}}
\put(0,10){\line(0,1){5}}
\put(5,15){\line(0,1){5}}
\put(10,20){\line(0,1){5}}
\put(15,25){\line(0,1){5}}
\put(20,30){\line(0,1){5}}
\put(25,35){\line(0,1){5}}
\put(30,40){\line(0,1){5}}
\put(35,45){\line(0,1){5}}
\put(40,50){\line(0,1){5}}
\put(10,5){\line(0,1){50}}
\put(25,5){\line(0,1){50}}
\put(-5,20){\line(1,0){50}}
\put(-5,35){\line(1,0){50}}
\put(16,0){$E_\alpha$} 
\put(-12,26){$E_\alpha$} 
\end{picture} 
\end{center}
\caption{The action of the number operator $n_\alpha$ on the joint Hilbert space $\mathfrak{H}^{(2)}$.  Blue squares represent fermion states and red squares represent boson states.  Light squares represent the $n_\alpha = 0$ eigenspace; medium squares represent the $n_\alpha = 1$ eigenspace; and dark squares  represent the $n_\alpha = 2$ eigenspace.  \label{NoOperator} }
\end{figure}
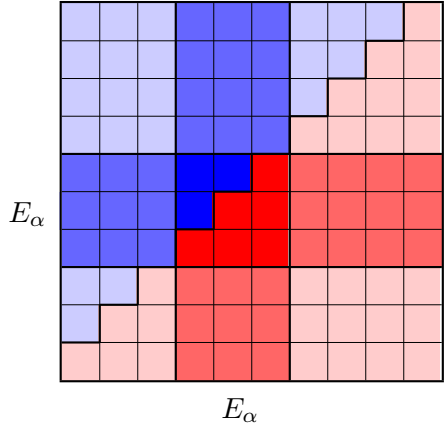

We begin with a chosen individuation criterion, a projector $E_\alpha$, which acts on the single-system Hilbert space $\mathcal{H}$.  Then it may be checked that the operator
\begin{equation}\label{nalpha}
n_\alpha := E_\alpha\otimes\mathds{1} + \mathds{1}\otimes E_\alpha
\end{equation}
is a number operator for the two-system assembly's Hilbert space. That is,  it ``counts" the number of systems which are picked out by $E_\alpha$.  It is represented in Figure \ref{NoOperator}.

We now define a linear map $\pi_\alpha : \mathcal{B}(\mathcal{H}) \to \mathcal{B}\left(\mathfrak{H}^{(2)}_\pm\right)$ from the single-system algebra into the joint algebra of the assembly, which takes single-system operators and gives the appropriate operator on the joint Hilbert space associated with the $\alpha$-system:
\begin{equation}\label{pialpha}
\pi_\alpha(Q) := E_\alpha QE_\alpha\otimes\mathds{1} + \mathds{1}\otimes E_\alpha QE_\alpha .
\end{equation}
(Note that, if $Q = E_\alpha QE_\alpha$, then $\pi_\alpha$ is just the \emph{symmetrizer} for $Q$.)  

I now claim that the expectation value for any single-system quantity $Q$, associated with ``the'' $\alpha$-system is given by
\begin{equation}\label{QIexpss}
\langle Q\rangle_\alpha := \frac{\langle \pi_\alpha (Q)\rangle}{\langle n_\alpha\rangle}\ .
\end{equation}
I will establish this claim by considering a few examples.
\begin{enumerate}
\item The state of the assembly $|\psi\rangle = \frac{1}{\sqrt{2}}\left(|\alpha\rangle\otimes|\beta\rangle \pm|\beta\rangle\otimes|\alpha\rangle\right)$, where $E_\alpha|\alpha\rangle = |\alpha\rangle$ and $E_\alpha|\beta\rangle=0$, and $Q|\alpha\rangle = q|\alpha\rangle$.  Then $\langle n_\alpha\rangle = 1$ and $\langle \pi_\alpha(Q)\rangle = q$; so $\langle Q\rangle_\alpha=q$.  That is, the system individuated by $E_\alpha$ takes as its expectation for $Q$ the value $q$, associated with the state $|\alpha\rangle$, for which individuation succeeds (i.e.,~the state that it is in the range of $E_\alpha$).  (Indeed, the $\alpha$-system is in an eigenstate for $Q$, since $\langle Q^2\rangle_\alpha = q^2$.)

\item  $|\psi\rangle = c_1\frac{1}{\sqrt{2}}\left(|\alpha_1\rangle\otimes|\beta_1\rangle \pm |\beta_1\rangle\otimes|\alpha_1\rangle\right) + c_2\frac{1}{\sqrt{2}}\left(|\alpha_2\rangle\otimes|\beta_2\rangle \pm |\beta_2\rangle\otimes|\alpha_2\rangle\right)$, where $|c_1|^2 + |c_2|^2=1$; and for all $i=1,2$: $E_\alpha |\alpha_i\rangle = |\alpha_i\rangle $ and $E_\alpha |\beta_i\rangle = 0$, and $Q|\alpha_i\rangle = q_i|\alpha_i\rangle$.  Then $\langle n_\alpha\rangle = 1$ and $\langle \pi_\alpha(Q)\rangle = |c_1|^2q_1 + |c_2|^2q_2$; so $\langle Q\rangle_\alpha=|c_1|^2q_1 + |c_2|^2q_2$.  That is, the system individuated by $E_\alpha$ takes as its expectation for $Q$ the average for all single-system states $|\alpha_i\rangle$, for which the individuation succeeds.  The weights for this average are given by the relative amplitudes of the non-GM-entangled terms.

\item $|\psi\rangle = |\alpha\rangle\otimes |\alpha\rangle$, for $|\alpha\rangle$ as above.  Then $\langle n_\alpha\rangle = 2$ and $\langle \pi_\alpha(Q)\rangle = 2q$; so $\langle Q\rangle_\alpha=q$.  In this case, $E_\alpha$ individuates \emph{two} systems, and the expectation (indeed, eigenvalue) for $Q$ for both of them is $q$.

\item  $|\psi\rangle = \frac{1}{\sqrt{2}}\left(|\alpha_1\rangle\otimes |\alpha_2\rangle \pm |\alpha_2\rangle\otimes |\alpha_1\rangle\right)$, for $|\alpha_1\rangle, |\alpha_2\rangle$ as above.  Then $\langle n_\alpha\rangle = 2$ and $\langle \pi_\alpha(Q)\rangle = q_1+q_2$; so $\langle Q\rangle_\alpha=\frac{1}{2}(q_1+q_2)$.   In this case, $E_\alpha$ again individuates two systems, one whose expectation value for $Q$ is $q_1$, and one whose value is $q_2$; thus we take the average.  However, the weights for this average are \emph{not} given, as above, by relative amplitudes for non-GM-entangled terms; (the entire state is non-GM-entangled).  Rather, they are given by the relative \emph{frequency}, in a single non-GM-entangled state, of each single-system state for which individuation succeeds.

\item  $|\psi\rangle = c_1U_\pm\left(|\alpha_1\rangle\otimes |\alpha_2\rangle \otimes |\beta_1\rangle\right) +  c_2U_\pm\left(|\alpha_3\rangle\otimes |\beta_1\rangle \otimes |\beta_2\rangle\right)$, where $U_\pm$ is the intertwiner which (anti-) symmetrizes product states, and the single-system states are defined as before. (So $N=3$; and for simplicity we set aside paraparticles.) Then $\langle n_\alpha\rangle = 2|c_1|^2 + |c_2|^2$ and $\langle \pi_\alpha(Q)\rangle = |c_1|^2(q_1+q_2) + |c_2|^2q_3$; so $\langle Q\rangle_\alpha=\frac{|c_1|^2(q_1+q_2) + |c_2|^2q_3}{2|c_1|^2 + |c_2|^2}$. In this case, the weights for the average are determined \emph{jointly} by relative amplitudes \emph{and} relative frequencies.  If $|c_1| = |c_2|$, then $\langle Q\rangle_\alpha=\frac{1}{3}(q_1+q_2 + q_3)$; thus each of the three states in the range of $E_\alpha$ are afforded equal weight, whether or not they belong to the same non-GM-entangled term. 

\end{enumerate}
Thus my claim---that the expectation value of any single-system quantity $Q$, for the system qualitatively individuated by $E_\alpha$, is given by (\ref{QIexpss})---yields the right results, at least for cases 1 to 4.  Case 5 seems to me less clear cut, since one might favour a different way to calculate statistical weights from the relative amplitudes and relative frequencies. However, I submit, there are no clear intuitions to rely on in this case, and I can see no objection to the way given by (\ref{QIexpss}).

The map $\pi_\alpha$ does \emph{not} constitute an isomorphism between the single-system algebra $\mathcal{B}(\mathcal{H})$---or indeed any subalgebra thereof---and the range of $\pi_\alpha$.  It is not even a homomorphism.  For example, it may be checked that $\pi_\alpha(AB) \neq \pi_\alpha(A)\pi_\alpha(B)$ does not hold, even for all those $A,B \in \mathcal{B}(\mathcal{H})$ that commute with $E_\alpha$.

This is not an objection to (\ref{QIexpss}), and should come as no surprise.  For there are states of the assembly in which $E_\alpha$ fails to individuate a \emph{unique} system (cf.~examples 3-5, above).  In these states, we should not expect that  $\pi_\alpha(AB) = \pi_\alpha(A)\pi_\alpha(B)$.  To perform the operation $B$, followed by $A$, on a given $\alpha$-system (corresponding to $\pi_\alpha(AB)$) relies on a \emph{re-identification} of that system (and that system alone) \emph{after} we have operated with $B$.  But the individuation criterion $E_\alpha$ cannot be guaranteed to pick out that very same system, if  more than one system is picked out by $E_\alpha$.\footnote{It is worth emphasising that the possibility of multiple ${{\alpha}}$-systems does {not} arise \emph{only} for bosons, since $E_\alpha$ need not be a  one-dimensional projector.  }   On the other hand, it may be checked that, for any two states $|\psi\rangle$ of the assembly that are eigenstates of $n_\alpha$ with eigenvalue 1---i.e.,~for all states in which exactly \emph{one} system is individuated by $E_\alpha$---we have $\langle\psi|\pi_\alpha(AB)|\psi\rangle = \langle\psi|\pi_\alpha(A)\pi_\alpha(B)|\psi\rangle$, as expected.

\subsection{Reduced density operators} \label{RDOs}

Thus we have a recipe for calculating the expectation value of any single-system quantity for a qualitatively individuated system or systems.  It remains for me to give a general prescription for calculating the reduced density operator for such a system.    We require that the reduced density operator $\rho_\alpha$ satisfy the condition that, for all $Q\in\mathcal{B}(\mathcal{H})$: $\mbox{tr}(\rho_\alpha Q) = \langle Q\rangle_\alpha$, as given in (\ref{QIexpss}) (I use the expression `tr' with a lowercase `t' to denote the trace over the single-system Hilbert space $\mathcal{H}$). We know from Gleason's Theorem that a unique such operator exists.

As usual, it is helpful to work by analogy with the case of ``distinguishable" systems.  The usual prescription for the reduced density operator of a constituent system, say the $k$th, of the assembly is (with $\rho$ the state of the assembly):
\begin{equation} \label{rdousual}
\rho_k := \mbox{Tr}_k\left(\rho\right),
\end{equation}
where Tr$_k$ denotes a partial trace over all but the $k$th factor Hilbert space.  Now this prescription is obviously no use to anti-factorists; but an equivalent formulation to (\ref{rdousual}) exists that will be of  far more use.  First we choose a complete orthobasis $\{|\phi_i\rangle\}$ for the single-system Hilbert space $\mathcal{H}$.  Then
\begin{equation}\label{rdousual2}
\rho_k := \sum^d_{i,j} \mbox{Tr}\left(\rho|\phi_j\rangle\langle\phi_i|_k\right) |\phi_i\rangle\langle\phi_j|
\end{equation}
where $d:=\mbox{dim}(\mathcal{H})$ and 
\begin{equation}\label{kthop}
|\phi_j\rangle\langle\phi_i|_k := \bigotimes^{k-1}\mathds{1}\otimes |\phi_j\rangle\langle\phi_i| \otimes \bigotimes^{N-k}\mathds{1}
\end{equation}
and we now perform a \emph{full} trace on the joint Hilbert space in (\ref{rdousual2}).

We may adapt (\ref{rdousual2}) for qualitatively individuated systems in the following way.  First, we replace each operator $|\phi_j\rangle\langle\phi_i|_k$, which is indexed to a factor Hilbert space, with $\pi_\alpha(|\phi_j\rangle\langle\phi_i|)$, as given in (\ref{pialpha}).  And second, we ``conditionalize" by dividing by $\langle n_\alpha\rangle = \mbox{Tr}(\rho n_\alpha)$.  Thus
\begin{equation} \label{RDO(i)}
\rho_\alpha = \frac{1}{\langle {n}_\alpha\rangle}\sum^d_{i,j} \mbox{Tr}\left[\rho\ \pi_\alpha(|\phi_j\rangle\langle\phi_i|)\right] |\phi_i\rangle\langle\phi_j| 
\end{equation}
Written out in full, and for any $N$, we have 
\begin{equation} \label{RDO(ii)}
\rho_{{\alpha}} =  \frac{\displaystyle\sum^d_{i,j} |\phi_i\rangle\langle \phi_j| \ \mbox{Tr}\left[\rho \left(\sum_{k=1}^N\bigotimes^{k-1}\mathds{1}\otimes E_\alpha| \phi_j\rangle\langle\phi_i|E_\alpha\otimes\bigotimes^{N-k}\mathds{1} \right)\right]}
{\displaystyle \mbox{Tr}\left[\rho \left(\sum_{k=1}^N\bigotimes^{k-1}\mathds{1}\otimes E_\alpha\otimes\bigotimes^{N-k}\mathds{1}\right)\right]}
\end{equation}
We may arrange for the first $d_\alpha := \mbox{dim}(E_\alpha)$ basis states to span $E_\alpha[\mathcal{H}]$; in which case
\begin{equation} \label{RDO(iii)}
\rho_{{\alpha}} =  \frac{\displaystyle\sum^{d_\alpha}_{i,j} |\phi_i\rangle\langle \phi_j| \ \mbox{Tr}\left[\rho \left(\sum_{k=1}^N\bigotimes^{k-1}\mathds{1}\otimes | \phi_j\rangle\langle\phi_i|\otimes\bigotimes^{N-k}\mathds{1} \right)\right]}
{\displaystyle \sum^{d_\alpha}_{i}\mbox{Tr}\left[\rho \left(\sum_{k=1}^N\bigotimes^{k-1}\mathds{1}\otimes  |\phi_i\rangle\langle \phi_i|\otimes\bigotimes^{N-k}\mathds{1}\right)\right]}
\end{equation}

It may then be shown (as required) that for any $Q\in\mathcal{B}(\mathcal{H})$, tr$(\rho_\alpha Q) = \langle Q\rangle_\alpha$.  For this, let $\{|\xi_i\rangle\}$ be a complete eigenbasis for $Q$, where $Q|\xi_i\rangle = q_i|\xi_i\rangle$.  Then
\begin{eqnarray}
\mbox{tr}(\rho_\alpha Q) &=& \frac{1}{\langle {n}_\alpha\rangle}\sum^d_{i,j,k} \mbox{Tr}\left[\rho\ \pi_\alpha(|\xi_j\rangle\langle\xi_i|)\right] \langle\xi_k|\xi_i\rangle\langle\xi_j|Q|\xi_k\rangle\\
&=& \frac{1}{\langle {n}_\alpha\rangle}\sum^d_{i,j,k} q_k\mbox{Tr}\left[\rho\ \pi_\alpha(|\xi_j\rangle\langle\xi_i|)\right] \delta_{ki}\delta_{jk}\\
&=& \frac{1}{\langle {n}_\alpha\rangle}\sum^d_{k} \mbox{Tr}\left[\rho\ q_k\ \pi_\alpha(|\xi_k\rangle\langle\xi_k|)\right] \\
&=& \frac{1}{\langle {n}_\alpha\rangle} \mbox{Tr}\left[\rho\ \pi_\alpha\left(\sum^d_{k} q_k\ |\xi_k\rangle\langle\xi_k|\right)\right]  \\
&=& \frac{1}{\langle {n}_\alpha\rangle} \mbox{Tr}\left(\rho\ \pi_\alpha\left(Q\right)\right) \\
&=:& \langle Q\rangle_\alpha .
\end{eqnarray}

Remember that  $\rho_{{\alpha}}$ as given in (\ref{RDO(i)}) and (\ref{RDO(ii)})  is the \emph{average} state of any system individuated by $E_\alpha$.  So long as the state $\rho$ is an eigenstate of ${n}_{{\alpha}}$ with eigenvalue 1---or even a superposition of ${n}_{{\alpha}} = 0$  and ${n}_{{\alpha}} = 1$ eigenstates---then $\rho_{{\alpha}}$ yields the state of \emph{the} ${\alpha}$-system.  However, if $\rho$ contains eigenstates with ${n}_{{\alpha}} > 1$, then the interpretation of $\rho_{{\alpha}}$ as the state of \emph{the} ${\alpha}$-system can no longer be sustained, since in those terms we effectively average over \emph{all} systems picked out by $E_{\alpha}$.

To conclude this section, I note two important limiting cases of Equation (\ref{RDO(i)}).   The first is when we are maximally discriminating in our individuation; i.e.,~where $E_\alpha$ is a one-dimensional projector.  Let $|\alpha\rangle$ be the state for which $E_\alpha|\alpha\rangle = |\alpha\rangle$.   Then, so long as $\langle n_\alpha\rangle >0$, $\rho_\alpha = |\alpha\rangle\langle\alpha|$, which is to be expected.  In this case the $\alpha$-system is a Fock space quantum, the object that is added to the field with the creation operator $a^\dag(\alpha)$.

The second limiting case lies at the other extreme, in which we individuate with maximum \emph{in}discriminateness, i.e.~with $E_{\bm{\alpha}} = \mathds{1}$.  In this case $\langle n_\alpha\rangle = N$ and $\pi_\alpha(A) = \sum_{k=1}^N\bigotimes^{k-1} \mathds{1}\otimes A\otimes\bigotimes^{N-k}\mathds{1}$; so 
\begin{eqnarray} \label{RDOE=1}
\rho_{{\alpha}} &=& \frac{1}{N}\sum^d_{i,j}|\phi_i\rangle\langle \phi_j| \ \mbox{Tr}\left[\rho\left( \sum_{k=1}^N \bigotimes^{k-1}\mathds{1}\otimes |\phi_j\rangle\langle \phi_i|\otimes\bigotimes^{N-k}\mathds{1}\right)\right] \\
&=& \frac{1}{N}\sum_{k=1}^N\sum^d_{i,j}|\phi_i\rangle\langle \phi_j| \ \mbox{Tr}\left[\rho\left(  \bigotimes^{k-1}\mathds{1}\otimes |\phi_j\rangle\langle \phi_i|\otimes\bigotimes^{N-k}\mathds{1}\right)\right] \\
&=& \frac{1}{N}\sum_{k=1}^N\sum^d_{i,j}|\phi_i\rangle\langle \phi_j| \ \mbox{Tr}\left[\rho\left(   |\phi_j\rangle\langle \phi_i|_k\right)\right] \quad \mbox{(from (\ref{kthop}))}\\
&=& \frac{1}{N}\sum_{k=1}^N\rho_k\qquad \mbox{(from (\ref{rdousual2}))} . \label{factrecover}
\end{eqnarray}
So with maximum indiscriminateness $\rho_\alpha$ is the ``average" of the standard reduced density operators obtained by partial tracing.  However, under PI we of course have $\rho_1 = \rho_2 = \ldots = \rho_k =:\overline{\rho}$, in which case $\rho_\alpha = \overline{\rho}$.  Thus we may say that, under PI, standard reduced density operators obtained by partial tracing codify only the state of the \emph{average system}, and not the state of any particular system.  In this sense, the factorist may be accused of committing the fallacy of misplaced concreteness, as when one takes the ``average taxpayer'' to be a real person.

The definition of $\overline{\rho}$ as the average reduced state obtained by partial trace also allows a more elegant expression for the reduced density operator $\rho_\alpha$ for the $\alpha$-system.  From (\ref{RDO(iii)}), we have
\begin{eqnarray} 
\rho_{{\alpha}} &=&  \frac{\displaystyle\sum^{d_\alpha}_{i,j} |\phi_i\rangle\langle \phi_j| \ \mbox{Tr}\left[\rho \left(\sum_{k=1}^N\bigotimes^{k-1}\mathds{1}\otimes | \phi_j\rangle\langle\phi_i|\otimes\bigotimes^{N-k}\mathds{1} \right)\right]}
{\displaystyle\sum^{d_\alpha}_{i} \mbox{Tr}\left[\rho \left(\sum_{k=1}^N\bigotimes^{k-1}\mathds{1}\otimes |\phi_i\rangle\langle\phi_i|\otimes\bigotimes^{N-k}\mathds{1}\right)\right]} \\
&=&  \frac{\displaystyle \sum^d_{i,j} E_\alpha|\phi_i\rangle\langle \phi_j|E_\alpha\ \mbox{Tr}\left[\rho \left(\sum_{k=1}^N\bigotimes^{k-1}\mathds{1}\otimes | \phi_j\rangle\langle\phi_i|\otimes\bigotimes^{N-k}\mathds{1} \right)\right]  }
{\displaystyle \mbox{tr}\left\{\sum^d_{i,j}E_\alpha|\phi_i\rangle\langle \phi_j|E_\alpha\ \mbox{Tr}\left[\rho \left(\sum_{k=1}^N\bigotimes^{k-1}\mathds{1}\otimes | \phi_j\rangle\langle\phi_i|\otimes\bigotimes^{N-k}\mathds{1} \right)\right] \right\}} \\\nonumber\\
&=& \frac{E_\alpha \overline{\rho}E_\alpha}{\mbox{tr}(\overline{\rho}E_\alpha)} \label{RDO(iv)} .
\end{eqnarray}
Thus $\rho_\alpha$ is obtained from the ``average state'' $\overline{\rho}$ by the usual L\"uder's rule, where we conditionalize on the single-system's state lying in $E_\alpha[\mathcal{H}]$.

\subsection{Qualitative individuation in Fock space} \label{Fock}

\emph{Expectation values and reduced density operators for Fock space.}
Fock space is just the direct sum of the $N$-system Hilbert spaces we have been separately considering:
 \begin{equation}
 \mathfrak{F}_{\pm}(\mathcal{H}) := \bigoplus_{N=0}^{\infty}
 \mathcal{S}^{(N)}_{\pm}\left(\bigotimes^N\mathcal{H}\right)
=:\bigoplus_{N=0}^{\infty}\mathfrak{H}^{(N)}_{\pm}
 \end{equation}
(where $\mathcal{S}_{\pm}^{(N)}$ is the $N$-system (anti-) symmetrization operator, where `$+$' corresponds to bosons, i.e.~symmetrization, and `$-$' to fermions, i.e.~anti-symmetrization).   The generalization to Fock space is straightforward, which vindicates the claim that anti-factorism about quantum mechanics meshes with quantum field theory in the limit of conserved total particle number.  We already have a prescription for calculating expectation values and reduced density operators for a system individuated by any criterion $E_\alpha$ in a joint Hilbert space of $N$ systems.  The strategy between the  Fock space generalizations of these prescriptions is simply to act separately according to them on subspaces of Fock space characterized by total particle number $N$.
 
First we define the ``number operator'' on Fock space as
\begin{eqnarray}
n_\alpha &:=& \bigoplus_{N=0}^\infty n^{(N)}_\alpha \\
&=&  \bigoplus_{N=0}^\infty \left(\left.\sum^N_{k=1}\bigotimes^{k-1}\mathds{1}\otimes E_\alpha\otimes\bigotimes^{N-k}\mathds{1} \right|_{\mathfrak{H}^{(N)}_{\pm}} \right) \label{NoOp0}
\end{eqnarray}

The appropriate function from the single-system algebra to the Fock space algebra $\pi^{\pm}_\alpha: \mathcal{B}(\mathcal{H})\to\mathcal{B}(\mathfrak{F}_{\pm}(\mathcal{H}))$ is then defined by its action on an arbitrary single-system operator $Q$:
\begin{equation}
\pi^{\pm}_\alpha(Q) =  \bigoplus_{N=0}^\infty \left(\left.\sum^N_{k=1}\bigotimes^{k-1}\mathds{1}\otimes E_\alpha QE_\alpha\otimes\bigotimes^{N-k}\mathds{1}\right|_{\mathfrak{H}^{(N)}_{\pm}} \right)
\end{equation}
and we have $\pi^{\pm}_\alpha(\mathds{1}) = n_\alpha$, as expected. and Expectation values for any single-system beable $A$ are given by $\langle Q\rangle_\alpha = \frac{\langle\pi^{\pm}_\alpha(Q)\rangle}{\langle n_\alpha\rangle}$. While the single-system expectation operator $\langle \cdot\rangle_\alpha: \mathcal{D(H)}\times\mathcal{B(H)}\to\mathbb{C}$ is as usual, now $\langle \cdot\rangle: \mathcal{D}(\mathfrak{F}_\pm(\mathcal{H}))\times\mathcal{B}(\mathfrak{F}_\pm(\mathcal{H})) \to \mathbb{C}$ is the expectation operation on density operators and bounded operators on Fock space.

\noindent\emph{Connections to the Fock space formalism.}
It will be instructive to connect the expressions above to  the more familiar aspects of the Fock space formalism, viz.~particle creation, annihilation and numbers operators.  Recall that we may define an \emph{annihilation} operator-valued function $a: \mathcal{H}\to\mathcal{B}(\mathfrak{F}_\pm(\mathcal{H}))$, which takes any single-system state $\phi\in\mathcal{H}$ to the annihilation operator  $a(\phi)$, defined by its action on an arbitrary state $\Psi$ of $\mathfrak{F}_{\pm}(\mathcal{H})$ (in tensor notation; see Geroch (2005, 29)):
\begin{equation}
\Psi = \xi_0 \oplus \xi_1\psi_{(1)}^a \oplus \xi_2\psi_{(2)}^{ab}\oplus\ldots \oplus \xi_N\psi_{(N)}^{a_1\ldots a_N}\oplus\ldots
\end{equation}
(where ${{\psi}_{(1)}}_a\psi_{(1)}^a={{\psi}_{(2)}}_{ab}\psi_{(2)}^{ab} = \ldots = {{\psi}_{(N)}}_{a_1\ldots a_N}\psi_{(N)}^{a_1\ldots a_N}= 1$; i.e.~each $N$-system state is normalized) as follows:
\begin{equation}
a(\phi)\Psi = \xi_1 \phi_n\psi_{(1)}^n\oplus \sqrt{2}\xi_2\phi_n\psi_{(2)}^{na} \oplus \ldots \oplus  \sqrt{N+1}\xi_{N+1}\phi_{n}\psi_{(N+1)}^{na_1\ldots a_N} \oplus\ldots
\end{equation}
And we similarly define the \emph{creation} operator-valued function $a^\dag: \mathcal{H}\to\mathcal{B}(\mathfrak{F}_\pm(\mathcal{H}))$, where the creation operator $a^\dag(\phi)$ is defined by its action on $\Psi$ as follows:
\begin{equation}
a^\dag(\phi)\Psi = 0 \oplus \xi_0\phi^a \oplus \sqrt{2}\xi_1\phi^{[a}\psi_{(1)}^{b]_{\pm}}\oplus  \ldots \oplus  \sqrt{N}\xi_{N-1}\phi^{[a_1}\psi_{(N-1)}^{a_2\ldots a_N]_{\pm}} \oplus\ldots
\end{equation}
Then the action of the number operator $\hat{N}(\phi) := a^\dag(\phi)a(\phi)$ is
\begin{equation}\label{NoOp}
\hat{N}(\phi)\Psi = 0 \oplus \xi_1\phi^a\phi_n\psi_{(1)}^n\oplus 2\xi_2\phi^{[a}\phi_n\psi_{(2)}^{|n|b]_{\pm}}\oplus \ldots \oplus  N\xi_N\phi^{[a_1}\phi_n\psi_{(N)}^{|n|a_2\ldots a_N]_{\pm}} \oplus\ldots
\end{equation}

These definitions, combined with the (anti-) symmetry of $\Psi$, entail the identity
\begin{equation} 
\hat{N}(\phi) \equiv \bigoplus_{N=0}^\infty \left(\left.\sum^N_{k=1}\bigotimes^{k-1}\mathds{1}\otimes |\phi\rangle\langle\phi| \otimes\bigotimes^{N-k}\mathds{1} \right|_{\mathfrak{H}^{(N)}_{\pm}} \right)
\label{NoOp2}
\end{equation}
(\emph{Proof sketch:} We consider separately the action of $\hat{N}(\phi)$ on each $N$-system component $\Psi|_{\mathfrak{H}^{(N)}_{\pm}}$ of an arbitrary state $\Psi\in\mathfrak{F}_\pm(\mathcal{H})$, and demonstrate the identity for each component. Consider for example the $N=2$ component.  From (\ref{NoOp}) we have $\hat{N}(\phi)\Psi|_{\mathfrak{H}^{(2)}_{\pm}} = 2\xi_2\phi^{[a}\phi_n\psi_{(2)}^{|n|b]_{\pm}} \equiv \xi_2(\phi^a\phi_n\psi_{(2)}^{nb} \pm \phi^b\phi_n\psi_{(2)}^{na})$. But due to the (anti-) symmetry of $\Psi$,  $\psi_{(2)}^{na} = \pm\psi_{(2)}^{an}$, so we have $\hat{N}(\phi)\Psi|_{\mathfrak{H}^{(2)}_{\pm}} =\xi_2(\phi^a\phi_n\psi_{(2)}^{nb} + \phi^b\phi_n\psi_{(2)}^{an}) \equiv \xi_2(\phi^a\phi_m\delta^b_n + \delta^a_m\phi^b\phi_n)\psi_{(2)}^{mn} \equiv \left.(|\phi\rangle\langle\phi|\otimes\mathds{1} + \mathds{1}\otimes|\phi\rangle\langle\phi|)\Psi\right|_{\mathfrak{H}^{(2)}_{\pm}}$, in agreement with (\ref{NoOp2}).)  It may also be checked that the usual (anti-) commutation relations hold, for all $\phi, \chi\in\mathcal{H}$:
\begin{equation}
[a(\phi), a(\chi)]_{\mp} = [a^\dag(\phi), a^\dag(\chi)]_{\mp} = 0; \qquad
[a(\phi), a^\dag(\chi)]_{\mp} = \phi_n\chi^n \equiv \langle\phi|\chi\rangle,
\end{equation}
where `$-$' now corresponds to bosons (commutation) and  '$+$'  to fermions (anti-commutation).

Now select some orthonormal basis $\{|\phi_i\rangle\}; i=1,\ldots, d_\alpha := \mbox{dim}(E_\alpha)$ that spans the space $E_\alpha[\mathcal{H}]$.  Then $E_\alpha =\sum_{i=1}^{d_\alpha}|\phi_i\rangle\langle\phi_i|$ and (combining (\ref{NoOp0}) and (\ref{NoOp2})),
\begin{equation}
n_\alpha = \sum_{i=1}^{d_\alpha} \hat{N}(\phi_i)
\end{equation}
This justifies our calling $n_\alpha$ a ``number operator'' in the first place---in fact $n_\alpha$ is just a generalisation of the usual Fock space number operators, in which we are typically less than maximally specific about the single-system state whose occupancy we are counting.  If we set $E_\alpha = |\alpha\rangle\langle\alpha|$ for some single-particle state $\alpha\in\mathcal{H}$ (the limit of maximum specificity), then $n_\alpha = \hat{N}(\alpha)$.
 If we set $E_\alpha = \mathds{1}$ (the limit of maximum \emph{non}-specificity), then $n_\alpha = \hat{N}$, the total number operator, defined by
\begin{equation}
\hat{N} := 0 \oplus \mathds{1}_\mathcal{H} \oplus 2\mathds{1}_{\mathfrak{H}^{(2)}_\pm} \oplus \ldots \oplus N \mathds{1}_{\mathfrak{H}^{(N)}_\pm} \oplus \ldots .
\end{equation} 
Thus $n_\alpha$ determines a spectrum of  number operators, reflecting the spectrum of specificity of our chosen individuation criterion $E_\alpha$, parameterized by its dimension $d_\alpha$, with the familiar state number operators and total number operator at opposite extremes.

The function $\pi^{\pm}_\alpha$ may be re-expressed as follows:
\begin{eqnarray}
\pi^{\pm}_\alpha(Q) &=&  \bigoplus_{N=0}^\infty \left[\left.\sum^N_{k=1}\bigotimes^{k-1}\mathds{1}\otimes \left(\sum_{i=1}^{d_\alpha}|\phi_i\rangle\langle\phi_i|\right)Q \left(\sum_{j=1}^{d_\alpha}|\phi_j\rangle\langle\phi_j|\right)\otimes\bigotimes^{N-k}\mathds{1}\right|_{\mathfrak{H}^{(N)}_{\pm}} \right] \\
&=&  \sum^{d_\alpha}_{i,j=1}\langle\phi_i|Q |\phi_j\rangle\bigoplus_{N=0}^\infty \left(\left.\sum^N_{k=1}\bigotimes^{k-1}\mathds{1}\otimes |\phi_i\rangle\langle\phi_j|\otimes\bigotimes^{N-k}\mathds{1}\right|_{\mathfrak{H}^{(N)}_{\pm}} \right) 
 \label{piFock}
\end{eqnarray}
We now make use of the identity 
\begin{equation}
a^\dag(\phi)a(\chi) \equiv \bigoplus_{N=0}^\infty \left(\left.\sum^N_{k=1}\bigotimes^{k-1}\mathds{1}\otimes |\phi\rangle\langle\chi|\otimes\bigotimes^{N-k}\mathds{1}\right|_{\mathfrak{H}^{(N)}_{\pm}} \right) 
\end{equation}
(which may be verified, as 
with $\hat{N}(\phi)$ above, by considering the action on each $N$-system component of an arbitrary Fock space state) to re-express (\ref{piFock}) as
\begin{equation}\label{piFock2}
\pi^{\pm}_\alpha(Q) = \sum_{i,j=1}^{d_\alpha} \langle\phi_i|Q |\phi_j\rangle\ a^\dag(\phi_i)a(\phi_j)
\end{equation}
And the general expression for the expectation value of $Q$ for the $\alpha$-system is
\begin{equation} \label{FockExp}
\langle Q\rangle_\alpha = \frac{\displaystyle\sum^{d_\alpha}_{i,j=1}\left\langle a^\dag(\phi_i)a(\phi_j) \right\rangle \langle\phi_i|Q|\phi_j\rangle}{\displaystyle\sum^{d_\alpha}_{i=1}\left\langle a^\dag(\phi_i)a(\phi_i) \right\rangle}
\end{equation}

If $Q$ has the $|\phi_i\rangle$ as eigenstates ($Q|\phi_i\rangle = q_i|\phi_i\rangle$)---which may always be arranged if $Q$ commutes with the individuation criterion; $[Q,E_\alpha] = 0$---then (\ref{FockExp}) simplifies to the reassuring expression
\begin{equation}
\langle Q\rangle_\alpha = \frac{\sum^{d_\alpha}_{i=1}\langle \hat{N}(\phi_i) \rangle q_i}{\sum^{d_\alpha}_{i=1}\langle \hat{N}(\phi_i) \rangle}
\end{equation}
which has a straightforward physical interpretation as being the average eigenvalue for $A$ of all Fock space quanta whose state satisfies the individuation criterion $E_\alpha$.
If the single-system operator and individuation criterion do not commute ($[Q,E_\alpha]\neq 0$), then the cross-terms in (\ref{piFock2}) survive, and will manifest experimentally as interference between the  single-system states $|\phi_i\rangle$.

A reduced density operator associated with the individuation criterion $E_\alpha$ can also be defined for Fock space, so that $\langle Q\rangle_\alpha = \mbox{Tr}(\rho_\alpha Q)$, for all $Q\in\mathcal{B}(\mathcal{H})$.  Given an orthonormal basis $\{|\phi_i\rangle\}$ for $\mathcal{H}$, we have (cf.~(\ref{RDO(i)}), above), 
\begin{equation}
\rho_\alpha = \sum^d_{i,j} |\phi_i\rangle\langle\phi_j|\left\langle |\phi_j\rangle\langle\phi_i|\right\rangle_\alpha
\end{equation} 
We can arrange for the first $d_\alpha$ basis states to span $E_\alpha[\mathcal{H}]$, in which case
\begin{equation}
\rho_\alpha = \frac{\displaystyle\sum^{d_\alpha}_{i,j}\left\langle a^\dag(\phi_i)a(\phi_j) \right\rangle|\phi_i\rangle\langle\phi_j|}
{\displaystyle\sum_{i=1}^{d_\alpha} \left\langle a^\dag(\phi_i)a(\phi_i) \right\rangle}
\end{equation} 
which is clearly a positive operator (since $\hat{N}(\phi_i)$ is, for all $\phi_i\in\mathcal{H}$) with unit trace.

\subsection{Ubiquitous and unique systems}

In section \ref{QIexp} above, I deliberately defined a procedure for calculating expectation values  for qualitatively individuated systems that does not depend on the joint state being an $n_\alpha = 1$ eigenstate.  This has advantages for purposes of generality, but  individuation criteria $E_\alpha$ for which the joint  state is an $n_\alpha=1$ eigenstate, and their associated systems, are also extremely useful.  This is because they are anti-factorist surrogates for ``distinguishable'' systems, which are identifiable in all states by the same factor Hilbert space label---and their individuation criteria are the anti-factorist surrogates for factor Hilbert space labels.   I will call such systems \emph{ubiquitous and unique}: \emph{ubiquitous}, because some single-particle state in $E_\alpha[\mathcal{H}]$ is occupied in each term of the joint state; \emph{unique}, because  such states occur only once in each such term.

Ubiquitous and unique systems, like ``distinguishable'' systems, probe the entire joint state, and never appear more than once; so their reduced states give valuable information about the joint state.  In particular, they will be helpful in the next section (\ref{EntCont}) in providing a means to define a continuous measure of entanglement for the joint state. 

If the $\alpha$-system is ubiquitous and unique, then we have by definition Tr$(\rho n_\alpha)= 1$ for the joint state $\rho$.  But $n_\alpha = N\Sigma^{(N)}(E_\alpha\otimes\mathds{1}\otimes\ldots\otimes\mathds{1})$, so
\begin{eqnarray}
\mbox{Tr}(\rho n_\alpha)\ =\ 1 &=& N\mbox{Tr}\left[\rho \Sigma^{(N)}\left(E_\alpha\otimes\bigotimes^{N-1}\mathds{1}\right)\right] \\
&=& \frac{1}{(N-1)!}\sum_{\pi\in S_N}\mbox{Tr}\left[\rho P(\pi)\left(E_\alpha\otimes\bigotimes^{N-1}\mathds{1}\right)P^\dag(\pi)\right] \\
&=& N\mbox{Tr}\left[\rho \left(E_\alpha\otimes\bigotimes^{N-1}\mathds{1}\right)\right] \\
&=& N\mbox{tr}\left(\rho_1 E_\alpha\right)  \\
&=& N\mbox{tr}\left(\overline{\rho} E_\alpha\right) 
\end{eqnarray}
where in the third step we use the cyclicity of the trace function and the fact that $P(\pi)\rho P^\dag(\pi) = \rho$, due to PI; in the fourth step we use the definition of the partial trace; and in the fifth step we use the fact that $\rho_1 = \overline{\rho}$ when PI is imposed.  Thus for ubiquitous and unique systems individuated by $E_\alpha$,
\begin{equation}
\mbox{tr}(\overline{\rho}E_\alpha) = \frac{1}{N}.
\end{equation}
and the expression (\ref{RDO(iv)}) for the reduced density operator for ``the'' $\alpha$-system simplifies to
\begin{equation}\label{RDO(v)}
\rho_\alpha = N E_\alpha\overline{\rho}E_\alpha .
\end{equation}
This has the following mathematical interpretation.  If $\overline{\rho}$ is written in matrix form, in a basis such that the first $d_\alpha$ states span $E_\alpha[\mathcal{H}]$, then the top-left block of this matrix---given by $E_\alpha\overline{\rho}E_\alpha$---is the matrix $\frac{1}{N}\rho_\alpha$.

Similar results follow for the $\beta$-system, and so on.  So
if a full $N$ ubiquitous and unique systems (with mutually orthogonal individuation criteria) can be found for the entire joint state---which will be the case iff the joint state lies in an individuation block---then $N\overline{\rho}$ can be put into a block-diagonal form with all of the qualitatively individuated systems' states as the diagonal entries:
\begin{equation}\label{RDO(vi)}
N\overline{\rho} = \left(
\begin{array}{ccc}
\rho_\alpha \\
&\rho_\beta \\
&& \ddots
\end{array}
\right)
\end{equation}
(where there are $N$ diagonal blocks). We see here quite vividly that $\overline{\rho}$ represents the ``average state'' of the constituent systems.  This result will come in useful in the next section.

\section{Entanglement} \label{Entangle}

In this section I draw out the implications of anti-factorism for the notion of entanglement.  First (section \ref{EntangleBetter}) I say why the traditional definition, in terms of non-separability, needs supplanting.  Then in section \ref{GMW}, I will connect these considerations with recent heterodox proposals for defining entanglement, or ``quantum correlations'', in the context of permutation invariance, and I present two results which secure the suitability of the agreed definition, and in section \ref{EntCont} offer a continuous measure for the notion; it will turn out to have a simple relation to the familiar von Neumann entropy of the reduced state.

\subsection{Against non-separability}\label{EntangleBetter}

If anti-factorism is right, then nothing in the formalism with physical significance can depend on the ordering of the factor Hilbert spaces in the joint Hilbert space.  But this immediately compels a distinction between two kinds of non-separability that a joint state may exhibit.

Separability, I take it, is a purely formal notion (at least, I will here take it to be): an $N$-system joint state $|\Psi\rangle$ is \emph{separable} iff there is a basis of the single-system Hilbert space $\mathcal{H}$ for which $|\Psi\rangle$ is some $N$-fold tensor product of those basis states. But a joint state may fail to be separable purely because it is a superposition of product states which differ only by how single-system states are distributed among the factor Hilbert spaces (e.g. $|\phi\rangle\otimes|\chi\rangle$ and $|\chi\rangle\otimes|\phi\rangle$).  Under the superselection rule induced by PI, the only such (pure) joint states are of the form
\begin{equation}\label{non-GMW}
|\psi^\pm(\phi)\rangle = \frac{1}{\sqrt{N!}}\sum_{\pi\in S_N} (\pm 1)^{\#(\pi)}P(\pi)|\phi(1)\rangle\otimes|\phi(2)\rangle\otimes\ldots\otimes|\phi(N)\rangle
\end{equation}
where $\phi: \{1,\ldots, N\}\to \{1,\ldots, d\}$ and $\{|1\rangle, \ldots, |d\rangle\}$  is an orthonormal basis for $\mathcal{H}$ (so $\phi$ may be thought of as a function that takes factor Hilbert space labels to single-system states).  In the case of bosons we allow $\phi$ not to be injective, but in that case the state in (\ref{non-GMW}) is not properly normalized.

If $\phi$ \emph{is} injective, then the joint state $|\psi^\pm(\phi)\rangle$ lies in an individuation block in $\mathfrak{H}^{(N)}_\pm$---in fact, it lies in a \emph{1-dimensional} individuation block.  The corresponding $N$  individuation criteria, which are 1-dimensional, are $E_i = |\phi(i)\rangle\langle\phi(i)|,\  i=1,2,\ldots N$.  Following our interpretative principle (above, section \ref{Problems2}) that intertwiners between unitarily equivalent reps preserve physical interpretations, we are compelled to give the same interpretation to $|\psi^\pm(\phi)\rangle$ that we give to each of the product states $P(\pi)|\phi(1)\rangle\otimes\ldots\otimes|\phi(N)\rangle$ when PI is not imposed.  But in these states, being separable, are unentangled.  It follows that, if entanglement is to be a \emph{physical}, rather than formal, concept, then we ought also to construe states of the form (\ref{non-GMW}) as unentangled.  (I have overlooked what to say about joint states of that form in which $\phi$ is not injective: this will be remedied in the following section.)

Some readers may baulk at this suggestion.  After all, the spin singlet state for two qubits,
\begin{equation}
\frac{1}{\sqrt{2}}(|\!\uparrow\rangle\otimes|\!\downarrow\rangle- |\!\downarrow\rangle\otimes|\!\uparrow\rangle),
\end{equation}
has the form (\ref{non-GMW}), yet this is the most commonly mentioned state in discussions of entanglement and in derivations of Bell-type inequalities (e.g.~Bell (1976))!  How could the paradigm of an entangled state \emph{not} count as entangled on a ``physical'' understanding of that concept?

The resolution to this little puzzle is simply that we need to be more careful about whether PI is or isn't being imposed.  In almost all textbook treatments, PI is not imposed and one derives a Bell inequality which the singlet state violates.  The key beables in that exercise are of the form $\sigma_i\otimes\mathds{1}$ or $\mathds{1}\otimes\sigma_j$ (and constructions out of these, chiefly $\sigma_i\otimes\sigma_j$), which manifestly make non-trivial use of the ordering of the factor Hilbert spaces, and therefore do not satisfy PI.  These beables belong to an algebra of operators defined on the \emph{full} joint Hilbert space $\mathfrak{H}^{(2)} = \mathbb{C}^4 \cong \mathbb{C}^2\otimes\mathbb{C}^2$.  In short: the constituent systems are treated as ``distinguishable''. 

If the systems are treated as ``indistinguishable''---i.e.,~when PI is imposed---then the singlet state constitutes the \emph{only} state accessible to two fermions; i.e.~$\mathfrak{H}^{(2)}_- = \mathbb{C}$.  The corresponding joint algebra is $\mathcal{B}(\mathbb{C})\cong\mathbb{C}$, i.e.~only $c$-numbers, which is far too meagre to accommodate the violation of a Bell inequality---it is too meagre even to define spin operators for each of the constituent systems (construed factoristically)!

In fact we know that the real-world systems whose spin-measurements violate Bell inequalities do not only possess spin as a degree of freedom. (This point and its implications are well put by Ladyman \emph{et al} (2013, 216).)  If we are pedantic about this, we ought also to include the spatial degrees of freedom. (After all, Stern-Gerlach apparatuses exist in space!)  Let us therefore consider the following \emph{four}-component state:
\begin{equation}\label{RealBell}
\frac{1}{{2}}(|L\rangle_1\otimes |R\rangle_2 + |R\rangle_1\otimes |L\rangle_2)\otimes(|\!\uparrow\rangle_1\otimes|\!\downarrow\rangle_2- |\!\downarrow\rangle_1\otimes|\!\uparrow\rangle_2),
\end{equation}
(where I have used labels to keep track of the single-system factor Hilbert spaces).
This state lies in the fermionic joint Hilbert space, and is \emph{not} of the form (\ref{non-GMW}), so it  counts as entangled even according to the current heterodox proposal.  Accordingly, it may be shown that it yields the same Bell-inequality-violating spin measurements in the context of PI as the singlet state does when PI is not imposed.

The equivalence is demonstrated by appealing to qualitative individuation.  We choose our individuation criteria for the two constituent systems to be $E_L := |L\rangle\langle L|\otimes\mathds{1}_{\mbox{\scriptsize{spin}}}$ and $E_R:=|R\rangle\langle R|\otimes\mathds{1}_{\mbox{\scriptsize{spin}}}$. Then it may be checked that, by (\ref{Op1}), the operators in the joint algebra corresponding to the spins (in units of $\frac{1}{2}\hbar$) of the systems qualitatively individuated by $E_L$ and $E_R$ are
\begin{eqnarray}
\sigma_i(L) := U_-(\mathbf{\sigma}_i\otimes\mathds{1}) U^\dag_- &=& (|L\rangle\langle L|\otimes \mathbf{\sigma}_i)\otimes(|R\rangle\langle R|\otimes\mathds{1}_{\mbox{\scriptsize{spin}}}) \nonumber\\
&&\qquad \qquad +\ (|R\rangle\langle R|\otimes\mathds{1}_{\mbox{\scriptsize{spin}}})\otimes(|L\rangle\langle L|\otimes \mathbf{\sigma}_i)
\\
\sigma_i(R) :=U_-(\mathds{1}\otimes \mathbf{\sigma}_i)U^\dag_- &=& (|R\rangle\langle R|\otimes \mathbf{\sigma}_i)\otimes(|L\rangle\langle L|\otimes\mathds{1}_{\mbox{\scriptsize{spin}}})\nonumber\\
&& \qquad \qquad +\ (|L\rangle\langle L|\otimes\mathds{1}_{\mbox{\scriptsize{spin}}})\otimes(|R\rangle\langle R|\otimes \mathbf{\sigma}_i)
\end{eqnarray}
which, given the state (\ref{RealBell}), yield the spin-correlation expectation values
\begin{equation}\label{spinco}
\langle u^i\sigma_i(L) v^j\sigma_j(R)\rangle = -u_iv^i
\end{equation}
for any two unit spatial vectors $u^i$ and $v^i$ (where summation over repeated indices is implied). This suffices to violate the Clauser-Horne-Shimony-Holt (CHSH) inequality
\begin{equation}\label{BELL}
\left|\langle u_1^i\sigma_i(L) v_1^j\sigma_j(R)\rangle + \langle u_1^i\sigma_i(L) v_2^j\sigma_j(R)\rangle + \langle u_2^i\sigma_i(L) v_1^j\sigma_j(R)\rangle - \langle u_2^i\sigma_i(L) v_2^j\sigma_j(R)\rangle\right| \leqslant {2}
\end{equation}
when suitable choices for $u_1^i, v_1^i, u_2^i, v_2^i$ are made (e.g.~in the $x$-$y$ plane: $u_1 = (0,1); u_2 = (1,0); v_1=\frac{1}{\sqrt{2}}(1,1); v_2 = \frac{1}{\sqrt{2}}(-1,1)$).

The singlet state, construed such that PI has \emph{not} been imposed, is frequently used as a simplifying surrogate for states of the form (\ref{RealBell}), for which PI \emph{has} been imposed.  (If your interest is in discussing entanglement, why complicate matters by worrying about permutation invariance?)  That the singlet does serve as a suitable simplifying surrogate is justified by appeal to the fact that it yields the spin-correlation expectation values
\begin{equation}
\langle u^i\sigma_i\otimes v^j\sigma_j\rangle = -u_iv^i ,
\end{equation}
in agreement with (\ref{spinco}).

It may also be of interest that the prescription (\ref{RDO(i)}) yields the following reduced density operators for the state (\ref{RealBell}): 
\begin{equation}
\rho_L = |L\rangle\langle L|\otimes \frac{1}{2}\mathds{1}_{\mbox{\scriptsize{spin}}}; \qquad
\rho_R = |R\rangle\langle R|\otimes \frac{1}{2}\mathds{1}_{\mbox{\scriptsize{spin}}}
\end{equation}
If we then trace out the spatial states, we obtain the density operator $\frac{1}{2}\mathds{1}_{\mbox{\scriptsize{spin}}}$ for each qualitatively individuated system---which is normally associated with the (factoristically construed!)~constituent systems of the singlet state when IP is not imposed.

Finally, I note that the state 
\begin{equation}
\frac{1}{\sqrt{2}}(|L\rangle_1\otimes|\!\uparrow\rangle_1\otimes |R\rangle_2\otimes |\!\downarrow\rangle_2 -  |R\rangle_1\otimes |\!\downarrow\rangle_1 \otimes|L\rangle_2\otimes|\!\uparrow\rangle_2)
\end{equation}
which \emph{is} of the form (\ref{non-GMW}), and therefore according to the current proposal ought not to count as entangled, does not violate the CHSH inequality (\ref{BELL}) for \emph{any} choice of $u_1^i, v_1^i, u_2^i, v_2^i$, and gives the reduced density operators $ \rho_L = |L\rangle\langle L|\otimes|\!\uparrow\rangle\langle\uparrow\!| , 
\rho_R = |R\rangle\langle R|\otimes|\!\downarrow\rangle\langle\downarrow\!|
$ (i.e.~\emph{pure} states) for the constituent systems: two features which, non-separability notwithstanding, we normally associate with the absence of entanglement.

\subsection{Connections to the physics literature I: GMW-entanglement}\label{GMW}

The above considerations are found in a series of papers by Ghirardi, Marinatto and Weber (Ghirardi, Marinatto \& Weber (2002); Ghirardi \& Marinatto (2003,2004,2005)).  They argue that the usual definition of entanglement (non-separability) can be naturally adapted to quantum systems governed by PI in the way just shown.  To avoid confusion with non-separability, I will call  their adapted definition, and the definition I wish here to endorse, \emph{GMW-entanglement}.

For simplicity, I assume $N=2$; generalizations for $N>2$ will be obvious.
Following Ghirardi, Marinatto and Weber (2002), define: 
\begin{quote}
The indistinguishable constituents of a two-system assembly are \emph{non-GMW-entangled} (we may say the joint state is non-GMW-entangled) iff \emph{both} systems have a complete set of properties.
\end{quote}
(The assembly is then defined as GMW-entangled iff it is not non-GMW-entangled.)  
 And we define:
\begin{quote}
Given a joint state $\rho$ of two ``indistinguishable'' systems, at least one  of the systems has a complete set of properties iff there exists a 1-dimensional projector  $E$, defined on ${\cal H}$, such that:
\begin{equation}
\label{wittgenstein}
  \mbox{Tr}(\rho \mathcal{E})=1
\end{equation}
where
\begin{equation}
  \label{operatorediproiezione}
   \mathcal{E} := E\otimes \mathds{1} + \mathds{1}\otimes E  - E\otimes E.
\end{equation}
\end{quote}
Because $E$ is 1-dimensional, Ghirardi and Marinatto say that at least one of the systems has a `complete set of properties'.  `Complete' here picks up on the fact that $E$ is minimal, and therefore maximally specific (maximally logically strong) about the system's state.  This almost corresponds above to the two systems being qualitatively individuated by the projectors $E$ and $E^\perp := (\mathds{1}-E)$.  The difference is that $\mathcal{E}$ also projects onto doubly-occupied joint states (which must be bosonic); thus we say `\emph{at least} one system has a complete set of properties'.

Ghirardi and Marinatto (2003, Theorems 4.2 \& 4.3) then prove the following theorem:
\begin{quote}
 At least one of the systems in a two-system assembly has a complete set of properties
 iff the assembly's state is obtained by symmetrizing or anti-symmetrizing a separable state.
\end{quote}
Here I give a quick sketch of their proof:\\
\emph{Right to left} (easy half):
 If $\left|\Psi\right\rangle $ is obtained by symmetrizing or
 anti-symmetrizing a factorized state of two indistinguishable constituents, then:
\begin{equation}
  \label{statononortogonale}
  |\Psi_\pm \rangle=\frac{1}{\sqrt {2(1\pm|\langle\phi|\chi\rangle|^2)}}\left(|\phi\rangle\otimes|\chi\rangle\pm
  |\chi\rangle\otimes|\phi\rangle\right).
\end{equation}
By expressing the state $|\chi\rangle$ as
\begin{equation}
 |\chi\rangle=\alpha |\phi\rangle +\beta|\phi^{\perp }\rangle,
  \qquad \langle\phi |\phi^{\perp }\rangle=0
 \label{Nb}
\end{equation}
and choosing $E=|\phi \rangle \langle\phi |$, one gets immediately
  \begin{equation}
   \mbox{Tr}(\rho \mathcal{E})\equiv \langle\Psi_\pm |\mathcal{E}|\Psi _\pm\rangle = 1.
   \label{Nc}
  \end{equation}

\noindent\emph{Left to Right} (hard half): If one chooses a complete orthonormal set of single-system states
 whose first element $|\phi_{0}\rangle :=|\phi\rangle$ is such that $E = |\phi_0\rangle\langle\phi_0|$,
 writing
\begin{equation}
  \label{statonorma}
  | \Psi_\pm  \rangle= \sum_{ij} c_{ij} \vert \phi_{i}\rangle
   \otimes \vert \phi_{j}\rangle, \qquad \sum_{ij} \vert
   c_{ij} \vert^{2}=1, \qquad c_{ji} = \pm c_{ij},
\end{equation}
and, using the explicit expression for $\mathcal{E}$ in terms of $E$, one obtains:
  \begin{equation}
  \label{pippo}
   \mathcal{E}|\Psi_\pm \rangle =
   |\phi_{0}\rangle \otimes \left( \sum_{j\neq 0}c_{0j}
   |\phi_{j}\rangle \right) +\left(\sum_{j\neq 0}c_{j0}
   |\phi_{j}\rangle \right) \otimes
   |\phi_{0}\rangle + c_{00}|\phi _{0}\rangle \otimes
   |\phi_{0}\rangle.
  \end{equation}
But now imposing condition (\ref{wittgenstein}) entails that $\mathcal{E}|\Psi_\pm\rangle = |\Psi_\pm\rangle$.

In the case of fermions, $c_{00}=0$.  So, introducing a normalized vector $\vert \xi\rangle :=
 \sqrt{2}\sum_{j\neq 0} c_{0j}|\phi_{j}\rangle$, we obtain
  \begin{equation}
  \label{fermioni}
   |\Psi_- \rangle = \frac{1}{\sqrt{2}} \left(\,
   |\phi_{0}\rangle\otimes|\xi\rangle -
   |\xi \rangle\otimes|\phi_{0}\rangle\right),
  \end{equation}
where $\langle\phi_{0}|\xi\rangle=0$.  For bosons, defining the following normalized vector
  \begin{equation}
   \vert \theta \rangle =\frac{1}{\sqrt{2- |c_{00}|^{2}}} \, \left(
   \sum_{j\neq 0} 2c_{0j} |\phi_{j} \rangle  + c_{00}
   |\phi_{0}\rangle \right),
  \end{equation}
the two-particle state vector  becomes
  \begin{equation}
  \label{duebosoni}
   \vert \Psi_+ \rangle = \frac{1}{2}\sqrt{2- |c_{00}|^2}\, \left(
   \vert \phi_{0}\rangle\otimes| \theta \rangle +
   \vert \theta \rangle\otimes| \phi_{0}\rangle\right).
  \end{equation}
Note that in this case the states $| \phi_{0}\rangle$ and
 $| \theta \rangle$ are orthogonal iff $c_{00}=0$, in which case $|\theta\rangle = |\xi\rangle$. $\Box$

Let us now consider GMW-entanglement specifically for fermions and bosons, respectively.

\emph{GMW-entanglement for fermions.}
Since $E\otimes E=0$ on $\mathfrak{H}^{(2)}_-$,
 one can drop such a term in all previous formulae.
Accordingly, $\mathcal{E}_{f}:=E\otimes \mathds{1}+\mathds{1}\otimes E$.

Due to the orthogonality of $\vert \phi_{0} \rangle$ and $\vert \xi \rangle$, for the state (\ref{fermioni}), we conclude not only that there is one fermion with a complete set of properties (given by $|\phi_{0} \rangle$) , but also that \emph{the other} fermion has a complete set of properties (given by  $| \xi \rangle$).

So according to the definition of GMW-entanglement, we have proved:
\begin{quote}
The fermions  of a two-system assembly, whose (pure) state is given by $|\Psi _-\rangle$,  are non-GM-entangled iff  $|\Psi_-\rangle$ is obtained by anti-symmetrizing a separable state.
\end{quote}

\emph{GMW-entanglement for bosons.}
The broad similarity to fermions is clear, especially from
 Equations (\ref{statononortogonale}) and (\ref{duebosoni}). As for fermions, the
 requirement that one of the two  bosons has a complete set of properties entails that the state is obtained by symmetrizing a separable state.  However, there are some remarkable differences from the fermion
  case. For bosons, three cases
 are possible, according to the single-system states that are the factors of the separable state:
\begin{enumerate}
\item  $| \theta\rangle \propto | \phi_{0} \rangle$, i.e.~$|c_{00}|=1$.
  Then the state is $| \Psi_+ \rangle = | \phi_{0}
   \rangle \otimes | \phi_{0} \rangle$ and one can infer that
  there are two bosons each with the \emph{same} complete set of properties given by  $E=\vert \phi_{0} \rangle \langle \phi_{0}\vert$. It may checked that for this state $\langle\Psi_+|E\otimes E|\Psi_+\rangle =1$.  $|\Psi\rangle$ belongs to \emph{no} individuation block.

\item  $\langle \theta| \phi_{0} \rangle =0$, i.e.~$c_{00}=0$. 
Then exactly the same argument as for fermions entails that one of the two bosons has a complete set of properties given by $E := |\phi_0\rangle\langle\phi_0|$ and the other of the two  bosons has a complete set of properties given by $F:=|\theta\rangle\langle\theta|$.  The joint state $|\Psi_+\rangle$ belongs the the 1-dimensional individuation block defined by these two single-system states.

\item  Finally, it can happen that $0< |\langle \theta | \phi_{0}\rangle| <1$. In this case, it is true from the definition above that there is \emph{a} boson with a complete set of properties given by $E$ and it is true that there is \emph{a} boson with a complete set of properties given by $F$.  But it is \emph{not true} that \emph{one} has a complete set of properties given by $E$ and \emph{the other} has a complete set of properties given by $F$.
For there is a non vanishing probability of finding both particles in the same state, since $\langle\Psi_+|E\otimes E|\Psi_+\rangle = \langle\Psi_+|F\otimes F|\Psi_+\rangle = |c_{00}|^2 >0$.
\end{enumerate}
According to our definition of GMW-entanglement, the joint state is non-GMW-entangled for the first two cases. But in the last case we cannot say that the joint state is non-GMW-entangled, even though we may say that one system has a complete set of properties given by $E$ and one system has a complete set of properties given by $F$.  The worry, of course, is that we are counting contributions from the same system each time, so we must resist the plural article `both'.  So any state of this third type counts as GMW-entangled, being a superposition of states of the first and second types. To sum up:

\begin{quote}
 \label{defofentangidentical3}
  The bosons of a two-system assembly, whose (pure) state is given by $|\Psi_+\rangle$, are non-GM-entangled iff either: (i) $|\Psi_+\rangle$ is obtained by  symmetrizing a factorized product of two orthogonal states; or (ii) $|\Psi_+\rangle$ is a product state of identical factors.
  \end{quote}

To connect with our usual terminology: a fermionic joint state is GMW-entangled iff it lies in a 1-dimensional individuation block.  A bosonic state is GMW-entangled iff: (i) it lies in a 1-dimensional individuation block (case 1, above); or (ii) it is a product state of identical factors.  For $N>2$, we must add a third clause: (iii) all sub-assemblies satisfy either (i) or (ii).  If the joint state lies in an individuation block, for both bosons and fermions the intertwiner between the individuation block and its corresponding product Hilbert space takes GMW-entanglement in the ``indistinguishable'' case to entanglement in the ``distinguishable'' case.

This is to be compared to the fact regarding ``distinguishable'' systems, that
their joint state is non-entangled iff the constituent systems (associated with factor Hilbert spaces) occupy {pure} states.   Both results have the same physical interpretation: that in non-(GMW-)entangled states, the joint state of the assembly supervenes on (i.e.~is uniquely determined by) the states of its constituent systems.  This is a good reason to take GMW-entanglement as the right notion of entanglement when PI is imposed.

A further reason is given by the following result concerning Bell inequalities.  From Gisin (1991), we know that, for two ``distinguishable'' systems, any joint state of theirs is entangled iff it violates a Bell inequality for some beables.  But we have seen that any 2-system individuation block is unitarily equivalent to some product Hilbert space for 2 systems, and that the intertwiner between them takes entanglement to GMW-entanglement.  Therefore we conclude:
\begin{quote}
 If the joint state lies in an individuation block, then:
{the joint state is GMW-entangled iff it violates a Bell inequality for some \emph{symmetric} beables.}  
\end{quote}

\subsection{Connections to the physics literature II: ``quantum correlations''}\label{QCs}

An alternative definition of entanglement for 2-system assemblies, the existence of \emph{quantum correlations}, has also been suggested by Schliemann \emph{et al} (2001) and Eckert \emph{et al} (2002).  They introduce the notion of the \emph{Slater rank} of a joint state (Schliemann \emph{et al} (2001, 8)), in analogy with the Schmidt rank of a joint state of two ``distinguishable'' systems.

For any such state, we can carry out a bi-orthogonal Schmidt decomposition:
\begin{equation}\label{Schmidt}
|\Psi\rangle = \sum_{i} c_{i}|\phi_i\rangle\otimes|\chi_i\rangle ,
\end{equation}
where $\langle\phi_i|\phi_j\rangle = \langle\chi_i|\chi_j\rangle =0$ for $i\neq j$.
The \emph{Schmidt rank} of $|\Psi\rangle$ is the minimum integer $r$, given all possible Schmidt decompositions, such that the sum (\ref{Schmidt}) has $r$ non-zero entries.  Clearly $|\Psi\rangle$ is entangled iff $r > 1$.

Now consider an arbitrary 2-fermion state,
\begin{equation}\label{Slater1}
|\Psi_-\rangle = \sum_{i<j} c_{ij}\frac{1}{\sqrt{2}}(|\phi_i\rangle\otimes|\phi_j\rangle - |\phi_j\rangle\otimes|\phi_i\rangle) 
\equiv  \sum_{i<j} c_{ij}a^\dag(\phi_i)a^\dag(\phi_j)|0\rangle ,
\end{equation}
where $|0\rangle$ is the Fock space vacuum, and $a^\dag(\phi_i)$ is the fermion creation operator for the state $|\phi_i\rangle$. What we now need is a fermionic analogue to the bi-orthogonal Schmidt decomposition.  This is supplied by the following result from matrix theory (see Schliemann \emph{et al} (2001, 3 (Lemma 1)) and Mehta (1977, Theorem 4.3.15)):

\begin{quote}
For any complex anti-symmetric $d\times d$ matrix $M$, there is a unitary transformation $U$ such that $\tilde{M} := UMU^T$ has the form
\begin{equation}
\tilde{M} = \mbox{diag}[Z_1, \ldots Z_r, Z_0]
\end{equation}
where
\begin{equation}
Z_k := 
\left(
\begin{array}{cc}
0 & z_k \\
-z_k & 0
\end{array}
\right), \ k \in\{1, \ldots, r\}
\quad
\mbox{and}
\quad
Z_0 := \mathds{O}^{(d-2r)}
\end{equation}
where $\mathds{O}^{(d-2r)}$ is the $(d-2r)\times(d-2r)$ null matrix.
\end{quote}
If we set  $M := c_{ij}$, then by an unitary transformation of the single-system Hilbert space $\mathcal{H}$, we can get the joint state $|\Psi_-\rangle$ in the form
\begin{equation}\label{Slater2}
|\Psi_-\rangle = \sum^{\llcorner\frac{1}{2}d\lrcorner}_{i=1} z_{i} a^\dag(\chi_{2i-1})a^\dag(\chi_{2i})|0\rangle .
\end{equation}
(where $\llcorner x\lrcorner$ is the largest integer no greater than $x$).
Now the \emph{fermionic Slater rank} of $|\Psi_-\rangle$ is the minimum number $r$ of non-zero terms in (\ref{Slater2}).  We may say that the joint state displays \emph{quantum correlations} iff $r>0$.

As Eckert \emph{et al} (2002, 12) remark, we do not see quantum correlations for dim$(\mathcal{H}) =: d<4$.  We can use our previous results to see give a physical insight into this claim.  First, we must see that any state of the form (\ref{Slater2}) lies in some individuation block in $\mathfrak{H}^{(2)}_-$.  There is some conventionality here, but the following choice of individuation criteria will do:
\begin{equation}
E_1 := \sum^{\llcorner\frac{1}{2}d\lrcorner}_i |\chi_{2i-1}\rangle\langle\chi_{2i-1}|;
\qquad
E_2 := \sum^{\llcorner\frac{1}{2}d\lrcorner}_i |\chi_{2i}\rangle\langle\chi_{2i}|.
\end{equation}
I.e., $E_1$ captures all odd states $|\chi_{2i-1}\rangle$ and $E_2$ captures all even states $|\chi_{2i}\rangle$. We may conclude that \emph{any 2-fermion state lies in an individuation block}; I will return to this in section \ref{FermionsAD}.

 Now the individuation block defined by $E_1$ and $E_2$ is unitarily equivalent to the joint Hilbert space $E_1[\mathcal{H}]\otimes E_2[\mathcal{H}]$.  From (\ref{Slater2}) we can see that the (inverse of the) intertwiner connecting these joint spaces sends $|\Psi_-\rangle$ to
\begin{equation}\label{Slater2}
U_-^\dag|\Psi_-\rangle = \sum^{\llcorner\frac{1}{2}d\lrcorner}_{i=1} z_{i} |\chi_{2i-1}\rangle\otimes|\chi_{2i}\rangle ,
\end{equation}
which has the bi-orthogonal Schmidt decomposed form of (\ref{Schmidt}).  \emph{Thus $|\Psi_-\rangle$ exhibits quantum correlations iff $U_-^\dag|\Psi_-\rangle$ is entangled.}
We also know that, for fermions, the intertwiner takes entanglement in the product space over to GMW-entanglement in the individuation block.  We conclude that \emph{any 2-fermion joint state exhibits quantum correlations (in the sense of Eckert \emph{et al} (2002)) iff it is GMW-entangled}.  

We can now see why, for two fermions to be exhibit quantum correlations, i.e.~be GMW-entangled, the single-system Hilbert space requires at least 4 dimensions.  For fermionic GMW-entanglement, the joint state must yield mixed states for both constituent systems.  This requires at least 2 states per system; the systems may not share any states, since their individuation criteria must be orthgonal.  This yields 4 single-system states in total.

Eckert \emph{et al} (2002) also offer a definition of \emph{bosonic quantum correlations}.  Consider an arbitrary 2-boson state,
\begin{equation}\label{Slater3}
|\Psi_+\rangle = \sum_{i<j} c_{ij}\frac{1}{\sqrt{2(1 + \delta_{ij})}}(|\phi_i\rangle\otimes|\phi_j\rangle + |\phi_j\rangle\otimes|\phi_i\rangle) 
\equiv  \sum_{i<j} c_{ij}b^\dag(\phi_i)b^\dag(\phi_j)|0\rangle ,
\end{equation}
where $|0\rangle$ is the Fock space vacuum, and $b^\dag(\phi_i)$ is the boson creation operator for the state $|\phi_i\rangle$. We now appeal to the following result from matrix theory (see Eckert \emph{et al} (2002, 9-10):
\begin{quote}
For any complex symmetric $d\times d$ matrix $M$, there is a unitary transformation $U$ such that ${M}' := UMU^T$ has the form
\begin{equation}
{M}' = \mbox{diag}[z_1, \ldots z_r, 0,\ldots, 0].
\end{equation}
\end{quote}
If we set $M:=c_{ij}$, then then by an unitary transformation of the single-system Hilbert space $\mathcal{H}$, we can get the joint state $|\Psi_+\rangle$ in the form of a superposition purely of doubly-occupied states:
\begin{equation}\label{Slater4}
|\Psi_+\rangle = \sum^{d}_{i=1} z_{i}|\chi_{i}\rangle\otimes|\chi_{i}\rangle .
\end{equation}
Then the \emph{bosonic Slater rank} is minimum number $r$ of doubly-occupied states with non-zero amplitudes.  \emph{Bosonic quantum correlations} correspond to $r>1$. 

It would be a mistake to follow Eckert \emph{et al} (2002, 24) in considering a bosonic Slater rank of more than 1 to clearly indicate anything like entanglement.  The best case for scepticism here is that the \emph{non}-GMW-entangled state.
\begin{equation}
|\Phi_+\rangle := \frac{1}{\sqrt{2}}(|\phi\rangle\otimes|\chi\rangle + |\chi\rangle\otimes|\phi\rangle)
\end{equation}
may be put into the form
\begin{equation}
|\Phi_+\rangle = \frac{1}{\sqrt{2}}(|\phi'\rangle\otimes|\phi'\rangle - |\chi'\rangle\otimes|\chi'\rangle)
\end{equation}
where
\begin{equation}
|\phi'\rangle := \frac{1}{\sqrt{2}}(|\phi\rangle+|\chi\rangle);
\qquad
|\chi'\rangle := \frac{1}{\sqrt{2}}(|\phi\rangle-|\chi\rangle);
\end{equation}
and so has a bosonic Slater rank of 2.  Thus we have agreement with Ghirardi \emph{et al} only for fermions.  In view of the unitary equivalence results above (section \ref{IndBlocks}), and the fact that intertwiners between individuation blocks and product spaces take GMW-entanglement to entanglement, we ought to retain the GMW definition of entanglement in the case of bosons.

\subsection{A continuous measure of entanglement for 2-system assemblies}\label{EntCont}

In the ``distinguishable'' case for assemblies of 2 systems (a.k.a.~``bipartite systems''), the entanglement of any pure joint state $|\Psi\rangle$ is commonly measured by the \emph{entropy of entanglement} $E(|\Psi\rangle)$, given by the von Neumann entropy of the reduced density operator of one of the consituent systems:
\begin{equation}
E(|\Psi\rangle) := S(\rho_1) = -\mbox{tr}({\rho_1} \log_2{\rho_1}) ,
\end{equation}
where $\rho_1$ is obtained by a partial trace, i.e.~$\rho_1 = \mbox{Tr}_1(|\Psi\rangle\langle\Psi|)$.
This measure meets the obvious requirement that it be independent of the choice of constituent system (i.e.~$E(|\Psi\rangle) = S(\rho_1) = S(\rho_2)$, where $\rho_2 = \mbox{Tr}_2(|\Psi\rangle\langle\Psi|)$), and it has the desirable properties that: (i) it takes its minimum value 0 for separable joint states, and (ii) it takes the maximum value $\log_2d$ when $\rho_1 = \frac{1}{d}\mathds{1}$, where $d := \mbox{dim}(\mathcal{H})$ (see Nielsen \& Chuang (2010, 513)). 

Due to the possibilities afforded by qualitative individuation, we can also define sensible reduced density operators under PI.  Thus we can use the von Neumann entropy of these reduced density operators to provide a new continuous measure of entanglement.  

We can use the unitary equivalence results above to export any fact holding  for  ``distinguishable'' systems over to ``indistinguishable'' systems, so long as the joint state lies in an individuation block.
In particular, we can export the fact that the von Neumann entropy of the reduced density operator of one of the constituents is equal to that of the other.  I.e.,~$S(\rho_\alpha) = S(\rho_\beta)$. (Note that it is important that systems be ubiquitous and unique.)  This fact, together with the form (\ref{RDO(vi)})  of $\overline{\rho}$  derived in the last section, we may infer
\begin{eqnarray}
S(\overline\rho) &:=& -\mbox{Tr}(\overline{\rho} \log_2\overline{\rho})  \\
&=& -\frac{1}{2}\mbox{Tr}[2\overline{\rho} \log_2(2\overline{\rho})]   +\log_22\\
&=& -\frac{1}{2}\left[\mbox{Tr}(\rho_\alpha \log_2 \rho_\alpha)
+
\mbox{Tr}(\rho_\beta \log_2 \rho_\beta)
\right] +1 \\
&=& \frac{1}{2}\left[S(\rho_\alpha)+ S(\rho_\beta)\right] +1 \\ 
&=& S(\rho_\alpha) + 1 .
\end{eqnarray}
Therefore the von Neumann entropy of any qualitatively individuated system differs from that of a factorist system only by a constant characterised by  the total number of systems, 2.

For fermion states, $S(\overline\rho)$ cannot be less that $\log_22$.  Under a factorist interpretation, this minimum indicates ``inaccessible'' entanglement brought about by the (anti-) symmetrization dictated by PI.  However, under an anti-factorist interpretation, this minimum is a result of Pauli exclusion, combined with the fact that $\overline{\rho}$ is not the state of any genuine system, but rather a statistical artefact, an ``average state''.  Given Pauli exclusion, this average state cannot be definite, and is less definite the more fermions there are; thus the entropy of the state cannot fall under $\log_22$.  On the other hand, the von Neumann entropies $S(\rho_\alpha)$ and $S(\rho_\beta)$ of the \emph{genuine} systems has a minimum value of zero, corresponding to a non-GMW-entangled joint state for the assembly.

The maximum value of $S(\overline\rho)$ is $\log_2 d$, where $d$ is the dimension of the single-system Hilbert space, corresponding to $\overline{\rho} = \frac{1}{d}\mathds{1}$.  It follows that, whenever the joint state lies in an individuation block, the qualitatively individuated systems cannot have a von Neumann entropy exceeding $\log_2(\frac{d}{2})$. In fact, the results of section \ref{QCs} entail that for fermions this upper limit can be refined to $\log_2(\llcorner\frac{d}{2}\lrcorner)$.  Finally, it may also be shown that the entropy of entanglement for a joint state of 2 fermions never exceeds $\log_2r$, where $r$ is its fermionic Slater rank.

All this suggests that a good measure of GMW-entanglement is given by $E_{\mbox{\scriptsize{\emph{GMW}}}}(|\Psi\rangle) := S(\overline{\rho}) - \log_22 \equiv E(|\Psi\rangle) - \log_22$.  As we have seen, this measure carries over to ``indistinguishable'' systems all of the desirable properties of the entropy of entanglement for ``distinguishable'' systems.  Its only drawback is that applies only when the joint state lies in an individuation block.  However, as we shall see in the next section, this is a restriction only for bosonic states.  How to measure entanglement for bosonic states which do not lie in an individuation block---i.e.~those that contain terms with identical factors---is still an open question.

\section{Implications for discernibility}\label{Fermions}

\subsection{When are ``indistinguishable'' quantum systems  discernible?}

As we saw in section \ref{IndBlocks},  any individuation block of either the  fermion or boson  $N$-system joint Hilbert space ``behaves'' in every way (restricted to that block), due to unitary equivalence, like a corresponding joint Hilbert space for ``distinguishable'' systems, in which PI is not imposed.  According to our interpretative principle regarding unitarily equivalent reps (\ref{Problems2}), we ought therefore to give the same physical interpretation to both joint Hilbert spaces.  Since in the ``distinguishable'' case, we are happy to say that our systems are discerned, since they always possess different properties (their corresponding single-system Hilbert spaces share no state in common), we ought to be happy to say the same in the case of our \emph{so-called} ``indistinguishable'' systems.
Therefore, if the assembly's state lies completely in an individuation block, the constituent systems are discernible. 

What kind of discernibility do we have here?  A modest tradition of taxonomizing grades of discernibility was inaugurated by Quine (1960), and has been continued by Saunders (2003b, 2006), Ketland (2011), Caulton \& Butterfield (2012a), Ladyman, Pettigrew and Linnebo (2012); and was crucial to identifying the subtle kind of discernbility in the results of Muller and Saunders (2008), Muller and Seevinck (2009), Caulton (2013) and Huggett and Norton (2013).  There is a general consensus that there are three main grades, here defined in model-theoretic terms.  We fix some language $\mathcal{L}$.  Then:
\begin{itemize}
\item \emph{Absolute discernibility.}  Two objects $a$ and $b$ are absolutely discernible in some model $\mathfrak{M}$ iff there is some monadic formula $\phi(x)$, expressible in $\mathcal{L}$ and not containing any singular terms,  such that $\mathfrak{M}\vDash \phi(a)$ and $\mathfrak{M}\vDash \neg\phi(b)$.
\item \emph{Relative discernibility.}  Two objects $a$ and $b$ are relatively discernible in some model $\mathfrak{M}$ iff there is some dyadic formula $\psi(x,y)$, expressible in $\mathcal{L}$ and not containing any singular terms,  such that $\mathfrak{M}\vDash \psi(a,b)$ and $\mathfrak{M}\vDash \neg\psi(b,a)$.
\item \emph{Weak discernibility.}  Two objects $a$ and $b$ are weakly discernible in some model $\mathfrak{M}$ iff there is some dyadic formula $\psi(x,y)$, expressible in $\mathcal{L}$ and not containing any singular terms,  such that $\mathfrak{M}\vDash \psi(a,b)$ and $\mathfrak{M}\vDash \neg\psi(a,a)$.
\end{itemize}
I say that there are three \emph{grades}, since it may be shown that absolute entails relative discernibility, and relative entails weak (Quine (1960, 230)).  Therefore absolute discernibility is the strongest kind.  And it is the kind exhibited by qualitatively individuated systems, since these systems (like their ``distinguishable'' counterparts) enjoy the physical interpretation of having orthogonal single-particle states, and the occupation of a single-particle state is expressible with a monadic predicate.

In summary: if the assembly's joint state lies in an individuation block, then its constituent systems are absolutely discernible.  So our question now is, In what cases does the assembly's joint state lie in some individuation block?

\subsection{Absolutely discerning fermions, and ``quantum counterpart theory''}\label{FermionsAD}

Here we use the results of sections \ref{GMW} and \ref{QCs}. In the discussion of fermions in section \ref{QCs} we saw that any 2-fermion joint state can be `Slater decomposed', and that individuation criteria may be chosen such that the joint state has support entirely within the individuation block so defined.  It follows that \emph{fermions are absolutely discernible in all states}.

It is worth emphasising that this result does not depend on the fermions in question possessing pure states; in fact the general form (\ref{Slater1}) of the 2-fermion joint state entails that the reduced density operators for the constituent fermions will typically be mixed.  So I am not here appealing to the undeniable fact that fermionic Fock space quanta---which \emph{are} always in pure states---are always absolutely discernible.  In fact, this latter result follows as special case of our general result for fermions, in which the individuation criteria are all 1-dimensional.

Recall too that there is \emph{conventionality} in the choice of individuation criteria which determine an individuation block that will capture the joint state.  Given the generic `Slater decomposed' form for the joint state,
\begin{eqnarray}
|\Psi_-\rangle &=& \sum^{\llcorner\frac{1}{2}d\lrcorner}_{i=1} z_{i} \frac{1}{\sqrt{2}}(|\chi_{2i-1}\rangle\otimes|\chi_{2i}\rangle - |\chi_{2i}\rangle\otimes|\chi_{2i-1}\rangle) \\
&\equiv&\sum^{\llcorner\frac{1}{2}d\lrcorner}_{i=1} z_{i}|\chi_{2i-1}\rangle\wedge|\chi_{2i}\rangle ,
\end{eqnarray}
(where we use for convenience the short-hand notation
\begin{equation}
|\phi\rangle\wedge|\chi\rangle 
:=
\frac{1}{\sqrt{2}}(|\phi\rangle\otimes|\chi\rangle - |\chi\rangle\otimes|\phi\rangle) ,
\end{equation}
which, in analogy with differential geometry, we might call the \emph{wedge product of $|\phi\rangle$ and $|\chi\rangle$}; I will say more about the  wedge product in Section \ref{PrefBasis}),
there are $2^{(\llcorner\frac{1}{2}d\lrcorner - 1)}$ different choices for the pair of individuation criteria, all of which yield a suitable individuation block, as illustrated in Figure \ref{TransState}.  

\begin{figure}[t]
\setlength{\unitlength}{1mm}  
\begin{picture}(140,70)   
  \definecolor{E1}{rgb}{.6, 0, .6}%
  \definecolor{E2}{rgb}{0, .6, .6}%
  \color{black} %
  \put(0,60){$|\Psi_-\rangle\ =$}
    \put(17,60){$z_1|\chi_1\rangle\wedge|\chi_2\rangle$}
    \put(38,60){$\ \ +\ z_2|\chi_3\rangle\wedge|\chi_4\rangle$}
        \put(71,60){$+ \quad\ldots \quad + \quad z_{\llcorner\frac{1}{2}d\lrcorner}|\chi_{2\llcorner\frac{1}{2}d\lrcorner - 1}\rangle\wedge|\chi_{2\llcorner\frac{1}{2}d\lrcorner }\rangle$}
        \color{E1}%
        \put(23,65){\line(0,1){3}}
        \put(23,68){\line(1,0){28}}
        \put(51,65){\line(0,1){3}}
        \put(52,65){\line(0,1){3}}
        \put(52,68){\line(1,0){24}}
         \put(79,68){\ldots}
          \put(108,65){\line(0,1){3}}
        \put(86,68){\line(1,0){22}}
        \put(60,70){$E_1$}
        \color{E2}%
           \put(34,54){\line(0,1){3}}
        \put(34,54){\line(1,0){27}}
        \put(61,54){\line(0,1){3}}
        \put(62,54){\line(0,1){3}}
        \put(62,54){\line(1,0){15}}
         \put(79,54){\ldots}
          \put(129,54){\line(0,1){3}}
        \put(86,54){\line(1,0){43}}
        \put(70,50){$E_2$}
        \color{black}%
          \put(0,30){$|\Psi_-\rangle\ =$}
    \put(17,30){$z_1|\chi_1\rangle\wedge|\chi_2\rangle$}
    \put(38,30){$\ \ +\  z_2|\chi_3\rangle\wedge|\chi_4\rangle$}
        \put(71,30){$+ \quad\ldots \quad + \quad z_{\llcorner\frac{1}{2}d\lrcorner}|\chi_{2\llcorner\frac{1}{2}d\lrcorner - 1}\rangle\wedge|\chi_{2\llcorner\frac{1}{2}d\lrcorner }\rangle$}
        \color{E1}
        \put(23,35){\line(0,1){3}}
        \put(23,38){\line(1,0){38}}
        \put(61,35){\line(0,1){3}}
        \put(62,35){\line(0,1){3}}
        \put(62,38){\line(1,0){14}}
         \put(79,38){\ldots}
          \put(108,35){\line(0,1){3}}
        \put(86,38){\line(1,0){22}}
        \put(60,40){$E_1$}
        \color{E2}
           \put(34,24){\line(0,1){3}}
        \put(34,24){\line(1,0){17}}
        \put(51,24){\line(0,1){3}}
        \put(52,24){\line(0,1){3}}
        \put(52,24){\line(1,0){25}}
         \put(79,24){\ldots}
          \put(129,24){\line(0,1){3}}
        \put(86,24){\line(1,0){43}}
        \put(70,20){$E_2$}
        \color{black}%
        \put(70,10){$\vdots$}
        \put(0,2){\emph{Choices on}}
        \put(0,-2){\emph{each branch:}}
        \put(27,0){1}
         \put(38,0){$\times$}
        \put(54,0){2}
         \put(70,0){$\times$}
          \put(79,0){\ldots}
        \put(88,0){$\times$}
        \put(105,0){2}
        \put(120,0){$=\quad 2^{(\llcorner\frac{1}{2}d\lrcorner - 1)}$}
\end{picture} 
\caption{Choices of individuation criteria for an arbitrary 2-fermion joint state.\label{TransState}}
\end{figure}

The physical interpretation of this is that there is no unique objective ``trans-state identity'' relation for qualitatively individuated systems.  As I remarked in section \ref{DecompRight}, this suggests a quantum analogue of Lewis's (1968) Counterpart Theory.  In that theory, intended to provide an alternative to orthodox modal logic, there is no unique, objective ``trans-world identity'' relation between individuals in different possible worlds; the result being that many modal claims depend for their truth on a (usually contextually determined---or under-determined!) choice of ``counterpart relations'' between individuals from different possible worlds.

Here, the lack of any unique, objective ``trans-state identity'' relation affects not only modal claims but also non-modal claims.  This is due to the possibility in quantum mechanics of superposing states to yield new states; so in particular, certain facts regarding a given GMW-entangled state hang on a choice of ``counterpart relations'' between the constituent systems in each of the non-GMW-entangled ``branches'' of which it is a superposition.   These counterpart relations are determined by a choice of individuation criteria, and the dependence of certain facts on this choice is reflected in the dependence on the choice of individuation criteria of  the effective joint state, as presceribed by (\ref{Strawson}).\footnote{Strictly speaking, an individuation criterion determines a counterpart relation that is more constrained than those permitted by Lewis.  Lewis's counterpart relations are deliberately designed so that they need not be equivalence relations; whereas  the relation `satisfies the same individuation criterion $E$ as' is manifestly an equivalence relation.}

The situation for bosons is partly given by the above results for fermions, together with the unitary equivalence results we have made so much use of. Specifically: for any GMW-entangled bosonic joint state lying in an individuation block, we can expect there to be an alternative choice of individuation criteria, and therefore an alternative individuation block, which captures the state.   Quantum counterpart theory goes for bosons as much as for fermions.

In fact in the bosonic case it may be argued that the situation is closer to Lewis's scheme: take for example the (GMW-entangled) state
\begin{equation}
|\Psi\rangle = \alpha|\phi\rangle\otimes|\phi\rangle + \beta\frac{1}{\sqrt{2}}(|\phi\rangle\otimes|\chi\rangle + |\chi\rangle\otimes|\phi\rangle)
\end{equation}
No individuation block captures this state, but it is sensible to extend our ideas of qualitative individuation and say that the two systems in $|\phi\rangle$ in the first ``branch'' (with amplitude $\alpha$) are \emph{both} counterparts to \emph{the} system in $|\phi\rangle$ in the second ``branch'' (with amplitude $\beta$).  Here the counterpart relation between branches determined by the individuation criterion $|\phi\rangle\langle\phi|$ fails to pick out a unique system in every branch, just as Lewis's counterpart relation is permitted to fail to pick out a unique individual in every world.

\subsection{Discerning bosons: whither weak discernibility?}\label{bosons}

As usual, anything said of fermionic joint states also applies to bosonic joint states that lie in some individuation block.  Therefore, both the absolute discernibility and the conventionality of ``trans-state identity'' carries over for bosons---so long as the joint state lies in an individuation block.

However,
 In section \ref{GMW}, we saw that there are bosonic states that lie in no individuation block: these states have non-zero amplitudes for multiply-occupied single-system states that cannot be removed under any change of basis in the single-system Hilbert space  (in Figure \ref{intertwiners}, this corresponds to non-zero amplitudes for square lying along the diagonal).  Our requirement that individuation criteria for distinct systems be orthogonal (one which the unitary equivalence result depends) means that these states can be captured by no individuation block.

Moreover, in the jargon of Ghirardi \& Marinatto (2003), we may say of doubly-occupied bosonic states that both bosons have a complete set of properties, as determined by some 1-dimensional projector.  Our unitary equivalence results do not apply here (since the joint state lies outside any individuation block), so there are no clues from there about how to interpret these states.  But, given that the number operator for some single-system state $|\phi\rangle$ has eigenvalue 2 in such a joint state, it is natural to interpret  this joint state as constituted by two systems, both in the state (represented by) $|\phi\rangle$.

Since the two bosons share the same state, we must conclude that they are not absolutely discernible.  It appears too that they must not be relatively discernible.  Might they yet be weakly discernible?  The claim that bosons \emph{are} weakly discernible has been defended recently (by Muller \& Seevinck (2009), Caulton (2013), Huggett \& Norton (2013)).  However, these claims depend for their physical significance on the assumption that factor Hilbert space labels represent or denote constituent systems, i.e.~they assume factorism.  

Take for example Theorem 2 of Caulton (2013):
\begin{quote}
For each state $\rho$ of an assembly of two particles, the relation $R'(\mathbf{Q},x,y)$ [defined by
\begin{equation}\label{Caulton2013}
R'(A,x,y) \quad\mbox{iff}\quad\frac{1}{4}\mbox{Tr}\left[\rho\left(A^{(x)} - A^{(y)}\right)^2\right] \neq 0
\end{equation}
where $A^{(1)} := A\otimes\mathds{1}$ and $A^{(2)}:=\mathds{1}\otimes A$] discerns particles 1 and 2 weakly \ldots where $\mathbf{Q}$ is the single-particle position operator, on the assumption of the Born rule. 
\end{quote}
The idea behind this theorem is that, \emph{even for product states with identical factors}, if the assembly comprises more than 1 constituent, then this is indicated by anti-correlations in \emph{some} single-system beable of which the constituent systems' states are not eigenstates.  These anti-correlations are taken to be measured by the (symmetric) beable $\Delta^2_A:=\frac{1}{4}\left(A^{(1)} - A^{(2)}\right)^2$. (We then take advantage of the fact that no eigenstates of $\mathbf{Q}$ exist in the single-system Hilbert space.)

However, despite that fact that $\frac{1}{4}\left(A^{(x)} - A^{(y)}\right)^2$, and therefore the relation $R'(A,x,y)$,  is permutation-invariant  for any $A, x,y$, the condition on the RHS of (\ref{Caulton2013}) may be interpreted as a relation between two particles \emph{only if} $x$ and $y$ take particle names as substitution instances.  In fact, they take factor Hilbert space labels as substitution instances (as indicated by the definitions of $A^{(x)}$), so it is only if factorism is right---which it is not---that the condition counts as a genuine relation between physical systems.  Thus the weak discernibility claim fails.

However, the theorem undoubtedly has \emph{some} physical interpretation.  Can't we still say of our two bosons in the joint state $|\phi\rangle\otimes|\phi\rangle$ that their states exhibit anti-correlations  in some basis?  Here we run into difficulty.
Under anti-factorism, the only way to refer to constituent systems is via individuation criteria---in this case we say that there are two bosons in the state $|\phi\rangle$.  Therefore the identity of the systems is inextricably tied to  our choice of individuation criteria: if we change the criteria, then we change the subject.  Therefore, to be able to talk about the sort of anti-correlations of which Caulton's (2013) Theorem 2 takes advantage, we would have to shift our individuation criteria to that basis, but by doing so we would thereby be talking about a \emph{different} collection of bosons.  Our original question, whether the two bosons in state $|\phi\rangle$ are weakly discernible, remains unanswered.

\subsection{Comparisons with classical particle mechanics}\label{CfClass}

At this point it may be illuminating to mark the similarities and differences between classical point-like particles and fermions and bosons.  Just as, in quantum mechanics, under PI we are led to consider the joint space of symmetric ($\mathfrak{H}^{(N)}_+$) or anti-symmetric ($\mathfrak{H}^{(N)}_-$) states, in classical mechanics, under PI, we are led to consider the reduced phase space (RPS) $\Gamma_{\mbox{\scriptsize{red}}}$, which is the standard phase space $\Gamma$---a Cartesian product of $N$ copies of the single particle phase space, $\Gamma = \prod^N \gamma$---quotiented by the symmetric group $S_N$ (see Belot 2001, 66-70).

The classical version of PI states that for all quantities $f$ with a physical interpretation, and all $\pi\in S_N$,
\begin{equation}
(f \circ P(\pi))(\xi) = f(\xi)
\end{equation}
for all joint states $\xi\in\Gamma$, where the symplectomorphism-valued function $P$ is defined by
\begin{equation}
P(\pi)(\langle q_1, p_1,  \ldots, q_n, p_n\rangle) := \langle q_{\pi^{-1}(1)}, p_{\pi^{-1}(1)},  \ldots, q_{\pi^{-1}(n)}, p_{\pi^{-1}(n)}\rangle .
\end{equation} 
Classical PI entails that all states in the same $S_N$-orbit give the same values for all physical quantities $f$, and it is this that justifies quotienting $\Gamma$ by $S_N$.

 Individuation blocks in the quantum case correspond to ``off-diagonal'' rectangular regions of the RPS; these regions have a Cartesian product structure, rather than the individuation blocks' tensor product structure, and they behave just like phase spaces for ``distinguishable'' particles (i.e. when PI is not imposed).

More specifically, let $E_i, i = 1, \ldots, N$ be $N$ disjoint regions of the single-particle phase space, i.e.~$E_i \subseteq \gamma$, and $E_i \cap E_j = \emptyset$ for all $i\neq j$.  Then the joint phase space $\Gamma_D := \prod^N_i E_i$ represents physical states in which all $N$ distinguishable particles necessarily have different positions or momenta.  We may now define a symplectomorphism $h: \Gamma_D \to R$ from this joint phase space onto a region $R \subset \Gamma_{\mbox{\scriptsize{red}}}$:
\begin{equation}
h (\xi) := \{(P(\pi))(\xi)\ |\ \pi\in S_N\} .
\end{equation}
$h$ maps points of $\Gamma_D$ to their orbits under $S_N$; these orbits  are the ``points'' of $R$.  $h$ induces a ``drag-along'' map $h^*$ on quantities on $\Gamma_D$; any quantity for ``distinguishable'' particles $f : \Gamma_D\to\mathbb{R}$ is sent to a permutation-invariant quantity $h^*f$ for ``indistinguishable'' particles, according to the rule $(h^*f) (\xi) := (f\circ h^{-1})(\xi)$, for all $\xi\in R$.  

This symplectomorphism warrants us to afford the same physical interpretation to the states of $R$ that we afford to the states of $\Gamma_D$; in particular, that the constituent particles are absolutely discernible, since they all possess different positions or momenta.  So long as the system point lies in $R$, we can adjudicate matters of trans-state (and therefore also trans-temporal) identity, talk about the same particles having different positions and momenta in different states, etc. But we are mindful that the existence of {different} product phase spaces $\Gamma'_D\subset\Gamma$ with  different symplectomorphisms $h'$ such that $h':\Gamma'_D\to R$ forces us to recognise a conventionality in these adjudications. (See Figure \ref{classical}).

\begin{figure}
\setlength{\unitlength}{1mm}  
\begin{picture}(150,60)   
 %
        \definecolor{grey}{rgb}{.8, .8, .8}
          \definecolor{fermions}{rgb}{.8, .8, 1}
        \definecolor{bosons}{rgb}{1, .8, .8}
          \definecolor{fermionsgrey}{rgb}{.4, .4, 1}
        \definecolor{bosonsgrey}{rgb}{1, .4, .4}
        \definecolor{fermiontext}{rgb}{0,0,.5}
        \definecolor{bosontext}{rgb}{.5,0,0}
\linethickness{15mm}
   \color{grey}
 \put(17.5,35){\line(0,1){20}}   
  \color{black}    
 \thicklines
\put(5,10){\framebox(50,50)}
  \linethickness{0.1mm}%
 \put(10,5){\line(0,1){50}}
 \put(25,5){\line(0,1){50}}
   \put(0,55){\line(1,0){25}}
  \put(0,35){\line(1,0){25}}
\thicklines%
\put(10,35){\framebox(15,20)}
\put(-2,21){$E_1$}\put(16,5){$E_1$}
\put(38,5){$E_2$} \put(-2,44){$E_2$} 
\put(27,0){\large $\gamma$}
\put(-10,32){\large $\gamma$}
\linethickness{15mm}%
\color{bosons}%
 \put(87.5,35){\line(0,1){20}} 
  \color{black}    %
 \thicklines%
\put(75,60){\line(1,0){50}}
\put(75,10){\line(0,1){50}}
\put(75,10){\line(1,1){50}}
 \thicklines%
\put(80,35){\framebox(15,20)}
\put(15,44){\Large $\Gamma_D$}
\color{bosontext}
\put(85,44){\Large $R$}
\color{black}
\thicklines%
\put(26,45){\vector(1,0){53}}
\put(63,47){$h$}
\end{picture} 
\caption{The symplectomorphism $h$ between the joint phase space spaces $\Gamma_D$ and $R$.  The square on the left represents the full phase space $\Gamma$ and the triangle on the right represents the full reduced phase space $\Gamma_{\mbox{\scriptsize{red}}}$. Cf.~Figure \ref{intertwiners}. \label{classical}}
\end{figure}

No single \emph{individuation region} (the term we may coin for regions such as $R$) can capture the entire RPS $\Gamma_{\mbox{\scriptsize{red}}}$, since $\Gamma_{\mbox{\scriptsize{red}}}$ does not have a global Cartesian product structure.  However, \emph{some} individuation region may be found for \emph{almost all} joint states: the only states that cause trouble are those in which two particles have exactly the same position and momentum.  (These states already cause trouble, since they lie on a boundary in the RPS, and therefore one cannot define tangent vectors for them.)

This far, the qualitative individuation of classical particles under PI is entirely analogous to that of their quantum counterparts.  This  is perhaps a surprise: for the quantum joint state, unlike the classical joint state, is not confined to a single point of the state space, but may rather be conceived as a wave that may spread throughout all of configuration space; yet, in the case of fermions and for most bosonic states, the state may still be confined to some individuation block.

These analogies should also bring reassurance to those wishing to conceptually reconcile rival approaches to imposing PI in quantum mechanics. This article has focussed exclusively on the procedure in which we first quantize and then impose PI; but one may instead proceed by first imposing PI on the classical joint phase space, and then quantizing.  The result is that (for 3 spatial dimensions or more) one automatically yields states belonging to either the boson or fermion sectors of $\mathfrak{H}^{(N)}$  (see Leinaas \& Myrheim (1977), Morandi (1992, ch.~3)).  We are therefore justified in thinking of the qualitative individuation present in this section \ref{QI} as the quantized version of qualitative individuation for classical particles on the RPS, as just outlined.

However, there are significant differences between the classical and quantum cases. I will discuss two here; they may both be considered a consequence of the possibility of  quantum superpositions and incompatible beables:
\begin{enumerate}
\item \emph{Disjunctive individuation.}  I remarked at the end of section \ref{IndBlocks} above that one cannot qualitatively individuate by using multiple individuation blocks at once, since this induces a superselection rule between individuation blocks and one therefore loses important information about the joint state.  The same is not true in the classical case: that is, one may mosaic the entire RPS---excluding the ``diagonal'' points---with individuation regions, and thereby capture every joint state that does not lie along the ``diagonal''.  (I call this ``disjunctive individuation'', since it has the logical form of a disjunction of conjuncts: e.g.~`either particle 1's state lies in $E^{(1)}_1$ while particle 2's state lies in $E^{(1)}_2$ \ldots, or particle 1's state lies in $E^{(2)}_1$ while particle 2's state lies in $E^{(2)}_2$ \ldots, or \ldots'.)

This also means that one may make the individuation regions as small as one likes without losing information.  This, in effect, is what is going on when we opt for a \emph{relational} means of discerning the two classical particles.  For example, for any 2-particle state in which the 1-particle configuration space is $\mathbb{R}$, the choice to label as particle ``1'' the particle that lies leftmost along the line (i.e.~$q_1 < q_2$) is equivalent to foliating the joint phase space with infinitely many individuation regions $R(q)$ (one for each value of $q$) in which the corresponding $\Gamma_D(q) := E_1(q)\times E_2(q) = \{\langle q_1,p_1;q_2, p_2 \rangle \ |\ q_1\leqslant q\ \&\ q_2> q\}$.

\item \emph{Incommensurable individuation criteria and composition.}  Another significantly novel aspect of qualitative individuation in quantum mechanics is that the same state may be captured by two different individuation blocks determined by mutually incompatible pairs of individuation criteria.  For example, the non-GMW entangled fermion state 
\begin{equation}
\frac{1}{\sqrt{2}}(|\phi\rangle\otimes|\chi\rangle - |\chi\rangle\otimes|\phi\rangle)
\end{equation}
may be captured by the 1-dimensional individuation block defined by the criteria
\begin{equation}
E^{(1)}_\alpha:=|\phi\rangle\langle\phi|; \qquad E^{(1)}_\beta:=|\chi\rangle\langle\chi|
\end{equation}
but this block is defined \emph{also} by the incompatible criteria
 \begin{equation}
E^{(2)}_\alpha:=|\phi'\rangle\langle\phi'|; \qquad E^{(2)}_\beta:=|\chi'\rangle\langle\chi'|
\end{equation}
 where
 \begin{equation}
 |\phi'\rangle := \alpha|\phi\rangle + \beta|\chi\rangle; \qquad
  |\chi'\rangle := -\beta^*|\phi\rangle + \alpha^*|\chi\rangle
 \end{equation}
 and $\alpha, \beta \neq 0$ or $1$.
 
 This phenomenon, which arises only for fermions but not bosons, has serious consequences for the sense in which an assembly is composed of its constituent systems.  This matter is pursued in AUTHOR (2014).
 \end{enumerate}

\section{Conclusion: why worry?} \label{Conclusion}

I have argued that we can take experimentally observed conformity with permutation-invariance in quantum mechanics as evidence for the conclusion that what is permuted in the formalism---the order of the factor Hilbert spaces, or the factor Hilbert space ``labels''---has no physical significance, just as in electromagnetism we take the gauge-invariance of all observed quantities as evidence against the physical significance of the four-vector potential.  The consequence of doing so is that some reform is necessary in the usual way of doing quantum mechanics. 

In particular, we need to reform our ideas of: (i) how to extract the state of any constituent system of an assembly; and (ii) what formal properties of the joint state indicate entanglement in a physically meaningful sense.  Accordingly, I have proposed: (i) a means of extracting states of constituent systems conducive to the demands of permutation-invariance, using a method I call `qualitative individuation'; and (ii) an alternative definition of entanglement, GMW-entanglement, for which non-separability is a necessary but not sufficient condition.  Both of these proposals rely on the fact that certain regions, called `individuation blocks', of the joint bosonic or fermionic Hilbert spaces are unitarily equivalent to joint Hilbert spaces associated with ``distinguishable'' systems, in which permutation-invariance is not imposed. 

The physical upshot of these proposals for the discernibility of ``indistinguishable'' systems are: (i) that fermions are always absolutely discernible; and (ii) that bosons are usually absolutely discernible, but sometimes utterly indiscernible.  A more unwelcome result is that joint states of fermions which, physically speaking, ought not to count as entangled, are aptly represented by \emph{subspaces} of the joint Hilbert space.  The physical interpretation of this result is that fermions cannot be said to compose their assemblies, at least in any normal sense of the word `compose'.

There is a tempting, simple response to the puzzles discussed above.  Since they arise from an attempt to understand quantum mechanics as a theory about objects that can be composed into assemblies, why not use the puzzles as a premise in an argument for the misguidedness of that understanding?

It can't be denied that this response engenders a certain relief at having cut the Gordian knot.  But on second thoughts, doubt and confusion creep back.  For an understanding of the single-system Hilbert space and its associated algebra of operators is normally our gateway to understanding the joint Hilbert space and \emph{its} associated algebra.  (It is what allows us, for example, to interpret the operator $(\sigma\otimes\mathds{1}+\mathds{1}\otimes \sigma)$ as the total spin operator.)  It hard to see that 
we can take these heuristics seriously once we have given up on the homely idea that an assembly is composed out of constituent systems.

However, there is an alternative physical picture, which gives prominence to the points or regions of space-time, and understands particles and their hitherto-composites as patterns in the properties assigned to these regions (see e.g.~Wallace \& Timpson 2010).  This picture is more familiar from quantum field theory.  It is noteworthy that we are led to it from considerations confined entirely to elementary quantum mechanics.

\section{References}
\ \\
\parindent=-12pt

Bell, J. S. (1976), `Einstein-Podolsky-Rosen experiments', \emph{Proceedings of the Symposium on Frontier Problems in High Energy Physics}, Pisa, pp.~33-45.

Belot, G. (1998), `Understanding Electromagnetism', \emph{British Journal for the Philosophy of Science} \textbf{49}, pp.~531-555.

Belot, G. (2001), `The Principle of Sufficient Reason', \emph{The Journal of Philosophy} \textbf{98}, pp.~55-74.

Butterfield, J.~N. (1993), `Interpretation and identity in quantum theory', \emph{Studies in the History and Philosophy of Science} \textbf{24}, pp. 443-76.

Butterfield, J. N. (2006), `The Rotating Discs Argument Defeated', \emph{The British Journal for the Philosophy of Science}, \textbf{57}, pp.~1-45.

Butterfield, J. N. (2011), `Against Pointillisme: A Call to Arms', in D. Dieks, W. J. Gonzalez, S. Hartmann, T. Uebel and M. Weber (eds.),\emph{Explanation, Prediction, and Confirmation: The Philosophy of Science in a European Perspective} Vol.~2, Dordrecht: Springer, pp.~347-365.

Caulton, A. (2013). `Discerning ``indistinguishable'' quantum systems', \emph{Philosophy of Science}
\textbf{80}, pp.~49-72.

Caulton, A. and Butterfield, J. N. (2012a), `On kinds of discernibility in logic and metaphysics', \emph{British Journal for the  Philosophy of Science} \textbf{63}, pp.~27-84. 

Caulton, A. and Butterfield, J. N. (2012b), `Symmetries and Paraparticles as a Motivation for Structuralism',  \emph{British Journal for the Philosophy of Science} \textbf{63}, pp.~233-285.

Dieks, D. and Lubberdink, A. (2011), `How Classical Particles Emerge from the Quantum World', \emph{Foundations of Physics} \textbf{41}, pp.~1051-1064.

Earman, J. (2004), `Laws, symmetry, and symmetry breaking: Invariance, conservation 
principles, and objectivity', \emph{Philosophy of Science} \textbf{71}, pp.~1227-1241. 

Eckert, K.,  Schliemann, J., Bru\ss, D. and Lewenstein, M. (2002), `Quantum Correlations in Systems of Indistinguishable Particles', \emph{Annals of Physics} \textbf{299}, pp.~88-127.

French, S. and Krause, D. (2006), \emph{Identity in Physics: A Historical, Philosophical and Formal Analysis.}  Oxford: Oxford University Press.

French, S. and Redhead, M. (1988), `Quantum physics and the identity of indiscernibles', \emph{British Journal for the Philosophy of Science} \textbf{39}, pp.~233-46.

Fuchs, C. (2002), `Quantum Mechanics as Quantum Information (and only a little more)', arXiv:quant- 
ph/0205039v1. Abridged version in \emph{Quantum Theory: Reconsideration of Foundations}, edited 
by A. Khrennikov (V\"axj\"o University Press, V\"axj\"o, Sweden, 2002), pp. 463-543.

Geroch, R. (2005), `Special topics in particle physics', unpublished lecture notes.  Available from http://www.strangebeautiful.com/other-texts/geroch-qft-lectures.pdf.

Ghirardi, G. and Marinatto, L. (2003), `Entanglement and Properties', \emph{Fortschritte der Physik}
\textbf{51}, pp.~379Ð387.

Ghirardi, G. and Marinatto, L. (2004), `General Criterion for the Entanglement of Two Indistinguishable Particles', \emph{Physical Review A} \textbf{70}, 012109-1-10.

Ghirardi, G. and Marinatto, L. (2005), `Identical Particles and Entanglement', \emph{Optics and Spectroscopy} \textbf{99}, pp.~386-390.

Ghirardi, G., Marinatto, L. and Weber, T. (2002),  `Entanglement and Properties of Composite Quantum Systems: A Conceptual and Mathematical Analysis', \emph{Journal of Statistical Physics} \textbf{108}, pp.~49-122.

Gisin, N. (1991), `Bell's Inequality Holds for all Non-Product States', \emph{Physics Letters A} \textbf{154}, pp.~201-202.

Greiner, W. and M\"uller, B. (1994), \emph{Quantum Mechanics: Symmetries}, 2nd revised edition.  Berlin: Springer.

Halvorson, H. (2007), `Algebraic Quantum Field Theory (with an appendix by Michael M\"uger)', in J. Butterfield, J. Earman (eds.), \emph{Philosophy of Physics (Handbook of the Philosophy of Science)} Part A (Amsterdam: North-Holland).

Healey, R. (2007), \emph{Gauging What's Real: the Conceptual Foundations of Contemporary 
Gauge Theories}. New York: Oxford University Press.

Healey, R. (2012), `Quantum theory: a pragmatist approach', \emph{British Journal for the Philosophy of Science} \textbf{63}, pp.~729-771.

Huggett, N. (1999), `On the significance of permutation symmetry', \emph{British Journal for the Philosophy of Science} \textbf{50}, pp.~325-47.

Huggett, N. (2003), `Quarticles and the Identity of Indiscernibles', in K. Brading and E. Castellani (eds.), \emph{Symmetries in Physics: New Reflections}, Cambridge: Cambridge University Press, pp.~239-249.

Huggett, N. and Imbo, T. (2009), `Indistinguishability', in D. Greenberger, K. Hentschel and F. Weinert (eds.), \emph{Compendium of Quantum Physics}, Berlin: Springer, pp.~311-317.

Huggett, N. and Norton, J. (2013), `Weak Discernibility for Quanta, the Right Way', forthcoming in the \emph{British Journal for the Philosophy of Science}.

Jordan, T. F. (1969), \emph{Linear Operators for Quantum Mechanics.} Duluth: Thomas F. Jordan.

Ketland, J. (2011), `Identity and Indiscernibility', \emph{Review of Symbolic Logic} \textbf{4}, pp.~171-185.

Ladyman, J., Linnebo, \O., and Bigaj, T. (2013), `Entanglement and non-factorizability', \emph{Studies in History and Philosophy of Modern Physics} \textbf{44}, pp.~215-221.

Ladyman, J., Linnebo, \O., and Pettigrew, P. (2012), `Identity and discernibility in philosophy and logic', \emph{Review of Symbolic Logic} \textbf{5}, pp.~162-86.

Leinaas, J. M. and Myrheim, J. (1977), `On the theory of identical particles', \emph{Il Nuovo Cimento} \textbf{37B}, p.~1-23.

Lewis, D. (1968), `Counterpart Theory and Quantified Modal Logic', \emph{Journal of Philosophy} \textbf{65}, pp.~113-126.

Lewis, D. (1986), \emph{On the Plurality of Worlds.}  Oxford: Blackwell.

Margenau, H. (1944), `The Exclusion Principle and its Philosophical Importance', 
\emph{Philosophy of Science} \textbf{11}, pp.~187-208. 

Massimi, M. (2001), `Exclusion Principle and the Identity of Indiscernibles: a Response to Margenau's Argument', \emph{The British Journal for the Philosophy of Science} \textbf{52}, pp.~303-330.

Messiah, A. M. L. and Greenberg, O. W. (1964), `Symmetrization Postulate and Its Experimental Foundation', \emph{Physical Review} \textbf{136}, pp.~B248-B267.

Merzbacher, E. (1997), \emph{Quantum Mechanics}, 3rd edition.  New York: Wiley.

Morandi, G. (1992), \emph{The Role of Topology in Classical and Quantum Physics}.  Berlin: Springer.  

Muller, F.~A. and Saunders, S. (2008), `Discerning Fermions', \emph{British Journal for the Philosophy of Science}, \textbf{59}, pp.~499-548.

Muller, F. A. and Seevinck, M. (2009), `Discerning Elementary Particles', \emph{Philosophy of Science}, \textbf{76}, pp.~179-200.

Murray, F. J. and von Neumann, J. (1936), `On rings of operators', \emph{Annals Of Mathematics} \textbf{37}, pp.~116-229.

Nielsen, M. A. and Chuang, I. L. (2010), \emph{Quantum Computation and Quantum Information}, 10th anniversary edition.  Cambridge: Cambridge University Press.

Pooley, O. (2006), `Points, Particles, and Structural Realism', in S. French, D. Rickles and J. Saatsi (eds.), \emph{The Structural Foundations of Quantum Gravity}, Oxford: Oxford University Press, pp.~83-120.

Prugove\v{c}ki, E. (1981), \emph{Quantum Mechanics in Hilbert Space}, 2nd edition. New York: Academic Press.

Quine, W. V. O. (1960), \emph{Word and Object.} Cambridge: Harvard University Press. 

Rae, A. I. M. (1992), \emph{Quantum Mechanics}, 3rd edition. Bristol: Institute of Physics.

Russell, B. (1905), `On Denoting', \emph{Mind} \textbf{14}, pp.~479-493.

Saunders, S. (2003a), `Physics and LeibnizÕs Principles', in K. Brading and E. Castellani 
(eds), \emph{Symmetries in Physics: Philosophical Reflections,} Cambridge: Cambridge 
University Press. 

Saunders, S. (2003b), `Indiscernibles, covariance and other symmetries: the case for non-reductive relationism', in A. Ashtkar, D. Howard, J. Renn, S. Sarkar and A. Shimony (eds.), \emph{Revisiting the Foundations of Relativistic Physics: Festschrift in Honour of John Stachel}, Amsterdam: Kluwer, pp.~289-307.

Saunders, S. (2006), `Are Quantum Particles Objects?', \emph{Analysis} \textbf{66}, pp.~52-63. 

Saunders, S. (2010), `Chance in the Everett interpretation', in Saunders, S., Barrett, J., Kent, A., and Wallace, D. (eds.),  \emph{Many Worlds?: Everett, Quantum Theory, \& Reality,} Oxford: Oxford University Press, pp.~181-205.

Schliemann, J., Cirac, J. I., Ku\'s, M., Lewenstein, M. \& Loss, D. (2001), `Quantum correlations in two-fermion systems.' \emph{Physical Review A} \textbf{64}: 022303.

Strawson, P. F. (1950),  `On Referring', \emph{Mind} \textbf{59}, pp.~320-344.

Tung, W.-K. (1985), \emph{Group Theory in Physics.}  River Edge: World Scientific.

Wallace, D. and Timpson, C. (2010), `Quantum Mechanics on Spacetime I: Spacetime State Realism', \emph{British Journal for the Philosophy of Science} \textbf{61}, pp.~697-727.

Weyl, H. (1928/1931), \emph{The Theory of Groups and Quantum Mechanics}. New York: Dover.

Zanardi, P. (2001), `Virtual Quantum Subsystems', \emph{Physical Review Letters} \textbf{87}, 077901-1-4.

Zanardi, P., Lidar, D. A. and Lloyd, S. (2004), `Quantum Tensor Product Structures are Observable-Induced', \emph{Physical Review Letters} \textbf{92}, 060402-1-4.

\end{document}